\documentclass{emulateapj} 
\usepackage{graphicx} 
\usepackage{epsfig}
\bibpunct{(}{)}{;}{a}{}{,} 
\usepackage{lscape}

\shorttitle{CASH II. A Sample of 16 Extremely Metal-poor Stars}  
 
\shortauthors{Hollek et al.} 
 
\begin{document} 

\title{The Chemical Abundances of Stars in the Halo (CASH)
  Project. II.  \\A Sample of 16 Extremely Metal-poor
  Stars\altaffilmark{1,2}}

\author{
Julie K. Hollek\altaffilmark{3}, 
Anna Frebel\altaffilmark{4}, 
Ian U. Roederer\altaffilmark{5}, 
Christopher Sneden\altaffilmark{3}, 
Matthew Shetrone\altaffilmark{3,6}, 
Timothy C. Beers\altaffilmark{7}, 
Sung-ju Kang\altaffilmark{8}, 
and Christopher Thom\altaffilmark{9}}

\altaffiltext{1}{Based on observations obtained with the Hobby-Eberly
Telescope, which is a joint project of the University of Texas at
Austin, the Pennsylvania State University, Stanford University,
Ludwig-Maximilians-Universit\"at M\"unchen, and
Georg-August-Universit\"at G\"ottingen.
}

\altaffiltext{2}{Based on observations gathered with the 6.5 meter
  Magellan Telescopes located at Las Campanas Observatory, Chile.}
\altaffiltext{3}{Department of Astronomy, University of Texas at
  Austin, Austin, TX 78712;
  julie,chris,shetrone@astro.as.utexas.edu}
\altaffiltext{4}{Harvard-Smithsonian Center for Astrophysics,
  Cambridge, MA 02138; afrebel@cfa.harvard.edu}
\altaffiltext{5}{Carnegie Observatories, Pasadena, CA 91101;
  iur@obs.carnegiescience.edu} 
\altaffiltext{6}{McDonald Observatory,
  University of Texas, Fort Davis, TX 78734}
\altaffiltext{7}{Department of Physics and
  Astronomy, and JINA: Joint Institute for Nuclear Astrophysics, Michigan
  State University, East Lansing, MI 48824; beers@pa.msu.edu}
\altaffiltext{8}{Department of Physics and Astronomy, Iowa State
  University, Ames, IA 50011; sjkang@iastate.edu}
\altaffiltext{9}{Space Telescope Science Institute, Baltimore, MD
  21218; cthom@stsci.edu}

\begin{abstract} 
We present a comprehensive abundance analysis of 20 elements for 16
new low-metallicity stars from the Chemical Abundances of Stars in the
Halo (CASH) project.  The abundances have been derived from both
Hobby-Eberly Telescope High Resolution Spectrograph snapshot spectra
(R $\sim15,000$) and corresponding high-resolution (R $\sim35,000$)
Magellan MIKE spectra.  The stars span a metallicity range from
$\mbox{[Fe/H]}$ from $-$2.9 to $-$3.9, including four new stars with
$\mbox{[Fe/H]}<-3.7$.  We find four stars to be carbon-enhanced
metal-poor (CEMP) stars, confirming the trend of increasing [C/Fe]
abundance ratios with decreasing metallicity.  Two of these objects can be
classified as CEMP-no stars, adding to the growing number of these
objects at [Fe/H]$<-3$.  We also find four neutron-capture enhanced
stars in the sample, one of which has [Eu/Fe] of 0.8 with clear
r-process signatures.  These pilot sample stars are the most
metal-poor ($\mbox{[Fe/H]}\lesssim-3.0$) of the brightest stars
included in CASH and are used to calibrate a newly-developed,
automated stellar parameter and abundance determination pipeline. This
code will be used for the entire $\sim500$ star CASH snapshot sample.
We find that the pipeline results are statistically identical for
snapshot spectra when compared to a traditional, manual analysis from
a high-resolution spectrum.
\end{abstract}

\keywords{Galaxy: halo---methods: data analysis---stars:
abundances---stars: atmospheres---stars: Population II}
 
\section{Introduction} 
The first stars formed from metal-free material in the early universe
and therefore are thought to have been massive ($\sim100M_{\odot}$;
e.g., \citealt{bromm99}).  Many of these first stars polluted the
surrounding local environment with their chemical feedback through
core-collapse supernovae.  From this enriched material, subsequent
generations of stars were born.  Due to the presence of additional
cooling mechanisms, these stars had a range of lower masses and thus
were longer lived (e.g.,~\citealt{brommnature}).

Today, we observe the surviving low-mass stars as the most metal-poor
stars in the Galaxy. The atmospheres of these objects contain the
chemical signatures of early supernova events.  By studying these
stars, constraints can be placed on the specific types of
nucleosynthetic events responsible for the observed abundances
patterns.

Efforts to classify metal-poor stars have been based upon metallicity,
$\mbox{[Fe/H]}$\footnote{\mbox{[A/B]}$ \equiv \log(N_{\rm A}/N_{\rm
    B}) - \log(N_{\rm A}/N_{\rm B})_\odot$ for N atoms of elements A,
  B, e.g., $\mbox{[Fe/H]}=-3.0$ is 1/1000 of the solar Fe abundance.}, and
chemical composition, $\mbox{[X/Fe]}$, to better understand the
diversity of the observed abundance patterns \citep{bc05}.  Stars with
$\mbox{[Fe/H]}<-3.0$ are labeled as extremely metal-poor (EMP).  The
metallicity distribution function shows that as metallicity decreases,
the number of stars in each metallicity bin rapidly decreases
\citep{Ryanetal:1991, CarneyMDF, Schoerck, li_mdf}.  Only $\sim25$ of
these EMP stars have $\mbox{[Fe/H]}\lesssim-3.5$.  Below
$\mbox{[Fe/H]} \sim-3.6$ there is a sharp drop in the number of stars,
so extreme EMPs are an important probe of this tail.  To date, only
three stars with $\mbox{[Fe/H]}<-$4.5 have been discovered, with two
considered to be hyper metal-poor ($\mbox{[Fe/H]}<-5.0$;
\citealt{HE0107_ApJ, HE1327Nature, he0557}).

The majority of metal-poor stars ([Fe/H]$< -1$) show abundance
patterns similar to the Solar System, but scaled down by metallicity,
with two main differences: there is an enhancement in the
$\alpha$-elements (e.g.,~[Mg/Fe]) and a depletion in some of the
Fe-peak elements (e.g.,~[Mn/Fe]) compared to the solar abundance
ratios.  This pattern can be explained with enrichment by previous
core-collapse supernovae (e.g.,~\citealt{heger_woosley2010}).  The
chemical outliers among stars with $\mbox{[Fe/H]}<-2.0$, which make up
perhaps $10\%$, show great diversity in their abundance patterns.
Many stars have overabundances in selected groups of elements, e.g.,
the rapid (r) neutron-capture process elements \citep{sneden_araa}
and/or the slow (s) neutron-capture process elements.  The frequency
of chemically unusual stars increases with decreasing metallicity,
with stars often belonging to multiple chemical outlier groups.  Not
included in this estimate of chemically unusual stars are the
so-called carbon-enhanced metal-poor (CEMP) stars (where
$\mbox{[C/Fe]}> 0.7$), which make up at least $\sim15\%$ of stars with
$\mbox{[Fe/H]}<-2.0$.  At the lowest metallicities, the frequency of
CEMP stars also increases
\citep{bc05,lucatello2006,frebel_bmps,cohen2006,carolloCEMP}.  In
fact, all three $\mbox{[Fe/H]}<-4.5$ stars are CEMP stars.

Medium-resolution spectra (R $\sim2,000$) can be used to
determine the overall metallicity based upon the strength of the Ca II K
line. However, medium-resolution spectra provide limited information
on the abundances of individual elements, especially at low
metallicity.  High-resolution (e.g., R $\sim40,000$) observations are
necessary to carry out detailed analyses which yield abundances with
small uncertainties ($\sim0.1$ dex).  These observations are, however,
more time consuming and require large telescopes to achieve an adequate
S/N ratio in the data.

``Snapshot'' spectra, with intermediate resolution (R
$\sim15,000$-$20,000$) and moderate S/N ($\sim40$) fill the gap
between time intensive high- and medium-resolution observations.  From
such snapshot data, abundances for $\sim15$ elements can be derived
with moderate uncertainties ($\sim0.25$ dex; \citealt{heresII}).  This
allows for a more efficient confirmation of EMP stars and chemical
outliers.  The \citet{heresII}, Hamburg/ESO R-process Enhanced Star,
(hereafter HERES) study itself determined abundances (and upper
limits) for a total of $\sim250$ stars based on VLT/UVES snapshot
spectra.

The Chemical Abundances of Stars in the Halo (CASH) project is a
dedicated effort that aims to provide abundances for $\sim500$
metal-poor stars primarily based on R $\sim15,000$, moderate-S/N
snapshot spectra taken with the High Resolution Spectrograph (HRS)
~\citep{tull98} on the fully queue scheduled \citep{HET} Hobby-Eberly
Telescope (HET).  One of the earliest results of the CASH project was
the discovery of an unusual, Li-enhanced giant, HKII~17435$-$00532
\citep{cash1}.  Given the large number of stars in the sample, it is
expected that there will be additional chemically-unusual stars.

In this paper, we study the CASH pilot sample of 16 stars, spanning a
metallicity range of $\sim1.0$ dex, from $\mbox{[Fe/H]} \sim-2.9$ to
$-3.9$.  The aim of the present study is twofold: to present abundance
analyses for the 16 most metal-poor stars included in the CASH project
and to use those abundances to calibrate the newly developed stellar
parameter and abundance pipeline.  We will use it to obtain abundances
for the full $\sim500$ star CASH sample (Hollek et al. 2012, in
prep.).

In Section~\ref{observations} we discuss the spectra in terms of the
sample selection, observational information, and data reduction.  In
Section~\ref{specanal} we introduce our spectral analysis tools,
including our linelist, equivalent width measurement routines, and
model atmosphere analysis code.  In Sections~\ref{stellarparams}
and~\ref{robustsp} we describe acquisition of our stellar parameters
for both sets of data and a comparison between the two.  In
Section~\ref{abundanal} we discuss the abundance analysis methods for
each element we measure, including the error analysis and comparison
of our results to those in the literature.  Section~\ref{interp}
includes a summary of our abundance results and discussion of the
implications of our derived abundances.  In Section~\ref{conclu} we
list our summary.

\section{Observations}\label{observations} 
\begin{figure*}[]
\includegraphics[width=1.0\textwidth]{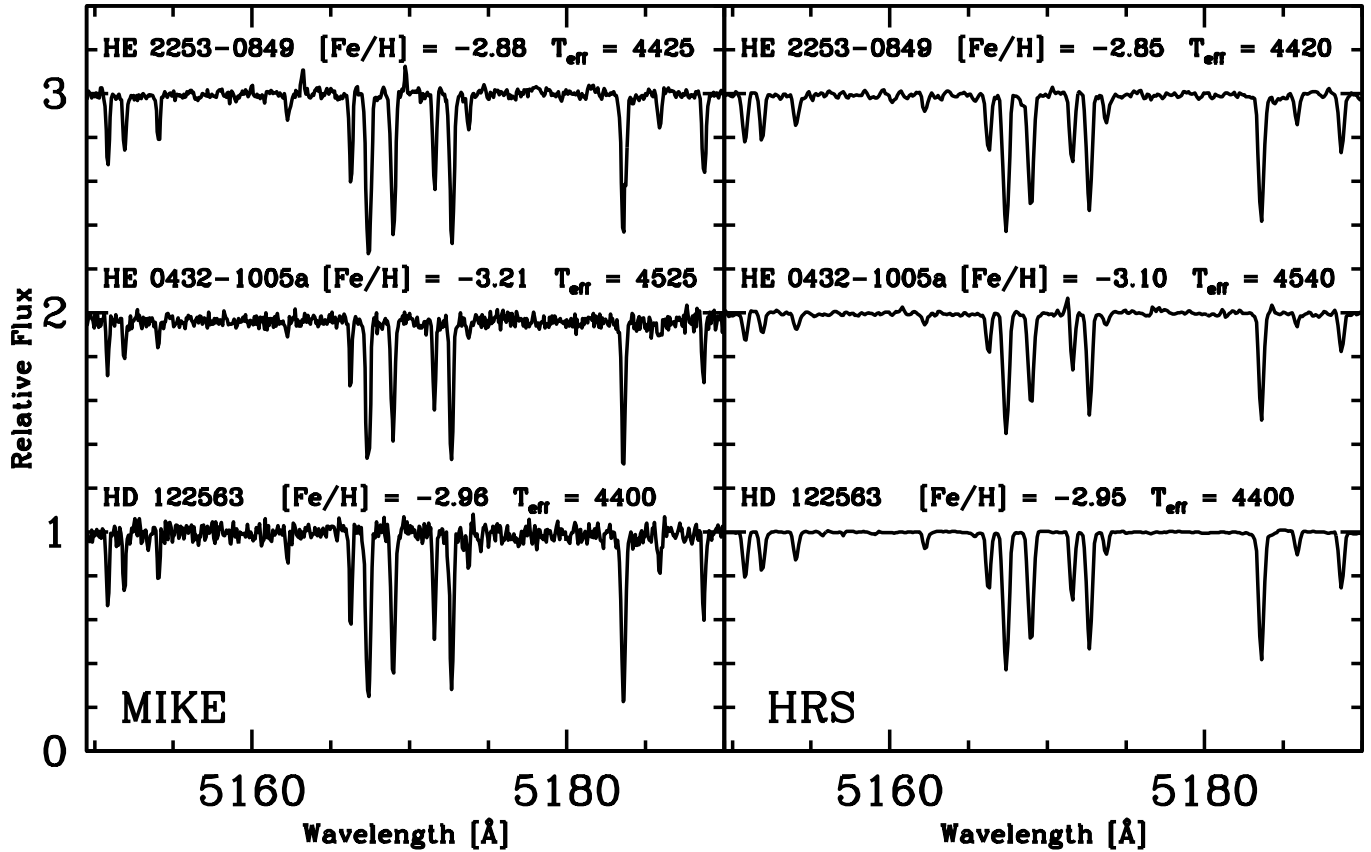}
\caption{\label{obs}MIKE (left) and HRS (left) spectra of three stars with
derived spectroscopic T$_{eff} \sim4500$ \,K. Both HRS and MIKE derived
[Fe/H] values are listed for each star. \label{mgspect}}
\end{figure*}

\subsection{Sample Selection} 
The stars of the pilot study were chosen from the Hamburg/ESO Bright
Metal-Poor Sample (BMPS; \citealt{frebel_bmps}) of the Hamburg/ESO
objective-prism plate Survey (HES; \citealt{stellar_content_I}). These
stars (B$<$14.5) required extra processing due to saturation of the
photographic plates.  Around 170 new metal-poor stars with
$\mbox{[Fe/H]}<-2.0$ were identified from medium-resolution spectra;
most of the BMPS stars observable from McDonald Observatory were added
to the CASH sample in order to obtain snapshot spectra with the HET.
This paper includes 16 HES BMPS objects, chosen as the most metal-poor
HES BMPS stars in the CASH study.  We also included five new
high-resolution observations of well studied stars from the literature
for comparison purposes.  All 16 HES objects have high-resolution
observations.  Of these objects, 14 have HRS spectra.  Only one
standard star, HD~122563, has an HRS spectrum.  Figure~\ref{mgspect}
shows both the HRS and MIKE spectra of three representative stars in
this sample.  The full $\sim500$ star sample contains targets from
other surveys including the Sloan Digital Sky Survey \citep{SDSS}, the
Sloan Extension for Galactic Understanding and Exploration
\citep{SEGUE}, and the HK \citep{HK1,HK2} surveys.

Of the HRS snapshot spectra included in this study, three spectra were
substantially contaminated with the solar spectrum, with no available
sky fibers to properly correct the data.  From these spectra, we were
able to determine that the [Fe/H] values for the stars were
roughly $-3.0$; however, these spectra require further processing in order
to derive accurate stellar parameters and abundances.  The
snapshot-derived results for these stars will be included in a future
paper, though we present the abundances derived from the
high-resolution MIKE spectra here.

\subsection{Spectroscopy}
The snapshot spectra for the CASH project were obtained using the
fiber-fed HRS on the HET at McDonald Observatory.  The CASH spectra
were obtained with a 2\mbox{\ensuremath{^{\prime\prime}}} fiber
yielding R $\sim15,000$.  The 2$\times$5 on-chip CCD binning leads to
3.2 pixels per resolution element. Two CCDs were used to record the
red and blue portions of the spectrum. The useful wavelength range is from
4200$-$7800\,{\AA}, or from the CH G-band to the oxygen triplet.  The
median S/N value for the entire 500 star sample is $\sim65$, with a
median S/N value for the pilot sample of 70; see Table~\ref{obs}.
There is substantially lower S/N at the blue ends of the spectra,
given the combination of the somewhat poor blue response of the HRS
and the lack of blue flux for many of the objects observed in CASH,
especially the cool giants.

High-resolution spectra for 21 stars were obtained using the MIKE
instrument \citep{mike} on the Magellan-Clay Telescope at Las Campanas
Observatory.  We used the 0\mbox{\ensuremath{^{\prime\prime}}}.7 slit
with 2$\times$2 on-chip binning, yielding a nominal resolution of R
$\sim35,000$ in the blue and 28,000 in the red with average S/N
$\sim85$ at 5200\,{\AA}.  MIKE spectra have nearly full optical
wavelength coverage from $\sim3500$-9000\,{\AA}. Table~\ref{obs} lists
the details of the observations for each star on both telescopes.

\begin{deluxetable*}{lccccrrr}
  \tablecolumns{8} 
  \tablewidth{0pc} 
  \tablecaption{Observations}
  \tablehead{ \colhead{Star}&\colhead{Telescope}&\colhead{UT Date}& \colhead{RA} &\colhead{Dec} &\colhead{t$_{exp}$} &\colhead{$S/N$} &\colhead{$v_{rad}$} \\
 \colhead{}&\colhead{}&\colhead{}& \colhead{(J2000)} &\colhead{(J2000)} &\colhead{sec} &\colhead{at 5180\,{\AA}} &\colhead{km~\,s$^{-1}$}  }

\startdata
              
HE~0013$-$0257 & HET      & 28 Jul 2007 & 00 16 04.2  &  $-$02 41 06  & 630  & 65    & 47.7   \\
               & Magellan & 28 Sep 2006 &             &               & 600  & 34    & 45.4  \\
HE~0013$-$0522 & HET      & 08 Aug 2007 & 00 16 28.1  &  $-$05 05 52  & 678  & 85    &$-$174.7\\
               & Magellan & 08 Aug 2010 &             &               & 1800 &81     &$-$175.5\\
HE~0015$+$0048 & HET      & 10 Aug 2007 & 00 18 01.4  &  +01 05 08    & 888  & 70    & $-$40.8\\
               & Magellan & 08 Aug 2010 &             &               & 1800 &56     &$-$48.8\\
HE~0302$-$3417a & Magellan & 27 Sep 2006 & 03 04 28.6  &  $-$34 06 06  & 300  &88     &121.7   \\
HE~0324$+$0152a & HET      & 24 Feb 2008 & 03 26 53.8  &  +02 02 28    & 316  &65     & 107.3  \\
               & Magellan & 27 Sep 2006 &             &               & 450  &70     & 106.1 \\
HE~0420$+$0123a & HET      & 05 Jan 2008 & 04 23 14.4  &  +01 30 49    & 207  & 180\tablenotemark{a}   & $-$53.3\\
               & HET      & 12 Nov 2009 &             &               & 800  &\nodata & $-$52.8\\
               & Magellan & 28 Sep 2006 &             &               & 300  & 64    &$-$55.3\\
HE~0432$-$1005a & HET      & 10 Nov 2008 &  04 35 01.2 &  $-$09 59 36  & 890  &80\tablenotemark{a}     & 198.0  \\
               & HET      & 12 Nov 2009 &             &               & 1800 &\nodata       & 199.6  \\
               & Magellan & 28 Sep 2006 &             &               &  900 & 44    &197.3  \\
HE~1116$-$0634 & HET      & contaminated\tablenotemark{b} & 11 18 35.8  &  $-$06 50 46  &\nodata&\nodata&\nodata\\
               & Magellan & 03 Jul 2010 &             &               &1200  & 121   & 115.5 \\
HE~1311$-$0131 & HET      & 06 Apr 2008 & 13 13 42.0  &  $-$01 47 16  & 250  &45     & 125.8  \\
               & Magellan & 05 Aug 2010 &             &               &2736  &48     & 124.7  \\
HE~1317$-$0407 & HET      & contaminated\tablenotemark{b} & 13 19 47.0  & $-$04 23 10   &\nodata&\nodata&\nodata \\
               & Magellan & 03 Jul 2010 &             &               & 487  & 135   &124.7 \\
HE~2123$-$0329 & HET      & 28 Jun 2008 & 21 26 08.9  &  $-$03 16 58  & 1473 &65     &$-$218.8\\
               & Magellan & 05 Aug 2010 &             &               & 1800 &85     &$-$219.4\\
HE~2138$-$0314 & HET      & 13 Nov 2009 & 21 40 41.5  &  $-$03 01 17  & 600  &85\tablenotemark{a}&$-$371.0\\
               & HET      & 14 Jul 2008 &             &               & 891  &\nodata&$-$371.6\\
               & Magellan & 05 Aug 2010 &             &               &1304  &93     &$-$373.5\\
HE~2148$-$1105a& HET      & contaminated\tablenotemark{b}& 21 50 41.5 &  $-$10 50 58  &\nodata &\nodata & \nodata\\
               & Magellan & 06 Aug 2010 &             &               & 300  & 61    &$-$87.17\\
HE~2238$-$0131 & HET      & 13 Nov 2008 & 22 40 38.1  &   $-$01 16 16 & 494  & 60    &$-$185.0\\
               & Magellan & 08 Aug 2010 &             &               & 1200 &96     &$-$186.7\\
HE~2253$-$0849 & HET      & 13 Nov 2008 &  22 55 43.1 &  $-$08 33 28  & 427  &75     & $-$89.4\\
               & Magellan & 08 Aug 2010 &             &               & 1200 &63     &$-$91.1\\
HE~2302$-$2154a & Magellan & 28 Sep 2006 & 23 05 25.2 &   $-$21 38 07 & 450  & 55    &$-$17.8\\
HD~122563      & HET      & 01 Mar 2009 & 14 02 31.8  & +09 41 10     & 30   &300    & $-$25.4\\
               & Magellan & 26 Jul 2009 &             &               & 5    &35     &$-$25.7 \\
CD~$-$38~245   & Magellan & 27 Jul 2009 & 00 46 36.2  & $-$37 39 33   & 250  & 65    & 47.1 \\
CS~22891$-$200 & Magellan & 05 Aug 2010 & 20 19 22.0  & $-$61 30 15   & 900  & 52    &137.7  \\
CS~22873$-$166 & Magellan & 27 Jul 2009 & 19 35 19.1  & $-$61 42 24   & 120  & 53    &$-$16.2\\
BD~$-$18~5550  & Magellan & 27 Jul 2009 & 19 58 49.7  & $-$18 12 11   & 87   & 157   &$-$125.3\\
\enddata 
\tablenotetext{a}{S/N for combined HET spectra}
\tablenotetext{b}{Sky contamination in spectrum, excluded from this analysis}
\label{obs}
\end{deluxetable*}

\subsection{Data Reduction} 

The HRS spectra were reduced using the IDL pipeline REDUCE
\citep{reduce}, which performs standard echelle reduction techniques
(trimming, bias subtraction, flat fielding, order tracing,
extraction).  The data were wavelength calibrated using ThAr lamp
exposures taken before or after every observation.  Comparisons have
been made between a by-hand IRAF\footnote{IRAF is distributed by the
  National Optical Astronomy Observatories, which is operated by the
  Association of Universities for Research in Astronomy, Inc., under
  cooperative agreement with the National Science Foundation.}
reduction and the REDUCE reduction of medium-S/N HRS data.  Both yield
comparable S/N across the spectrum, and the measured equivalent widths
for 121 different lines differ between the two different reductions by
3$\pm$8m\,{\AA}, which is statistically insignificant \citep{cash1}.  In
addition, earlier tests of REDUCE versus IRAF have shown that the optimized
extraction in REDUCE for high S/N spectra yields an extracted spectrum
that is less noisy than that of a spectrum extracted in IRAF (see
Figure 8 in \citealt{reduce}).  Standard IRAF routines were then used
to coadd (in the case of multiple observations) and continuum
normalize the individual observations into a final one-dimensional
spectrum.  Radial velocities (RVs) were computed by cross-correlating
the echelle order containing the Mg b triplet against another
metal-poor giant observed with the same instrumental setup.  Typical
uncertainties were 2$-$3 km/s for a single observation.  Barycentric
velocity corrections were computed using the IRAF ``rvcorrect'' routine.

Spectra observed using the MIKE instrument were reduced using an
echelle data reduction pipeline made for MIKE\footnote{available at
  http://obs.carnegiescience.edu/Code/python} and then normalized and
coadded using the same method as the HRS spectra.

\section{Spectral Analysis}\label{specanal}
\subsection{Linelist}x
The linelist to analyze the HRS spectra was based on the lines
included in \citet{roedererlinelist}.  Only those lines which are
unblended at the median S/N and resolution of the typical HRS snapshot
spectrum were included in our final list.

The linelist for the MIKE data is a composite of the lines from
\citet{roedererlinelist}, supplemented with additional lines from
\citet{cayrel2004}, and \citet{aoki_studiesIV}.  This linelist
includes those lines used for the HRS snapshot spectra analysis.  In
the instances where the same line was included in more than one
linelist, the most up to date oscillator strength was used, following
\citet{roedererlinelist}.  We confirmed that all important lines for
our abundance analysis were included in this linelist by plotting the
position of each line against a high-resolution MIKE spectrum of a
star with a higher metallicity than that of any star in the pilot
sample.  Compared to the sample stars, this star displays many more
absorption lines. This also allowed us to visually inspect for
features that were not present in both an EMP star and star that is
still considered to be metal-poor, but with substantially higher (1
dex) [Fe/H].

\subsection{Line Measurements}
In each spectrum, we measured equivalent widths of unblended lines of
various elements.  Table~\ref{eqwtab} lists the element and ionization
state, equivalent widths, wavelength, excitation potential, and
oscillator strengths of each measured line in the MIKE spectra.  These
equivalent widths were used to determine stellar parameters and
abundances for $\sim10$ elements.  The equivalent widths in the HRS
spectra were measured using an IDL routine.  Here we briefly summarize
the features important to this work.  The routine works to
automatically fit a Voigt profile to each line.  The user can then
manually adjust the continuum level, the number of spectral points
over which the line is fit, and the line center, among other features.
Given the limitations in resolution and S/N, some lines used in the
analysis for HRS spectra depart from the linear portion of the curve
of growth, thus line fits using the Voigt profile, rather than simply
a Gaussian fit, are preferred.

\begin{deluxetable}{llcrrr}
  \tablecolumns{6}
  \tablewidth{0pc}
  \tablecaption{Equivalent Widths\label{eqwtab}}
  \tablehead{\colhead{Star}&\colhead{Ion}&\colhead{Wavelength}&\colhead{XP}&  \colhead{log gf} &\colhead{EW}\\
            \colhead{}&\colhead{}&\colhead{\,{\AA}}&\colhead{}&  \colhead{} &\colhead{m\,{\AA}}}
\startdata
HE~0015+0048 & 12.0 & 3986.75 & 4.35 & $-$1.030 & 27.5\\
HE~0015+0048 & 12.0 & 4057.50 & 4.35 & $-$0.890 & 41.9\\
HE~0015+0048 & 12.0 & 4167.27 & 4.35 & $-$0.710 & 64.9\\
HE~0015+0048 & 12.0 & 4571.09 & 0.00 & $-$5.688 & 57.7\\
HE~0015+0048 & 12.0 & 4702.99 & 4.33 & $-$0.380 & 65.1\\
HE~0015+0048 & 12.0 & 5528.40 & 4.34 & $-$0.498 & 78.2\\
HE~0015+0048 & 13.0 & 3961.52 & 0.01 & $-$0.340 &125.1\\
HE~0015+0048 & 20.0 & 4226.73 & 0.00 &    0.244 &210.3\\
HE~0015+0048 & 20.0 & 4289.36 & 1.88 & $-$0.300 & 51.7\\
HE~0015+0048 & 20.0 & 4318.65 & 1.89 & $-$0.210 & 42.5\\
\ldots\\
\enddata

\tablecomments{Table~\ref{eqwtab} is published in its entirety in the
  electronic edition.  A portion is shown here for guidance regarding
  its form and content.  We list the ionization state of each element
  where .0 indicates a neutral species and .1 indicates a
  singly-ionized species.}
\end{deluxetable}

The equivalent widths in the high-resolution spectra were measured
with an ESO/Midas routine which automatically fits Gaussian profiles to
each line.  The user can calculate a fit to the continuum level by
selecting line free continuum regions.  This code takes into account
any possible non-zero slope of the continuum.

We chose a different equivalent width measurement routine because the
higher S/N of the MIKE spectra allowed us to detect deviations from
zero in the slope of the continuum that may arise from small-scale
variation in imperfectly normalized spectra or nearby strong lines.
Thus, it was helpful to be able to make a linear fit when determining
the continuum.  Additionally, the larger wavelength range allowed for
more lines to be measured, thereby enabling us to exclude lines near the
flat part of the curve of growth.  Table~\ref{eqwtab} lists the
equivalent widths for all our stars in the pilot sample.

\begin{figure}[]
\includegraphics[width=0.5\textwidth]{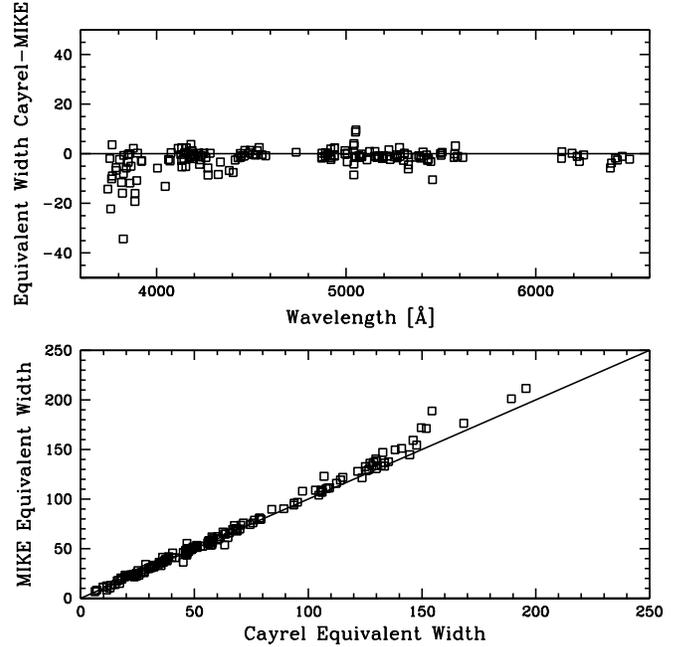}
\caption{Comparison of the equivalent widths measured
  from the MIKE and \citet{cayrel2004} spectra of BD~$-$18 5550 .  In
  the upper panel, residuals of the equivalent widths (Cayrel$-$MIKE)
  are plotted against wavelength.  In the bottom panel, the measured
  equivalent widths from Cayrel and MIKE are plotted against each
  other.}
\label{bdcayca}
\end{figure}

Figure~\ref{bdcayca} shows our measured equivalent widths from the
MIKE spectrum of BD~$-$18~5550 plotted against the equivalent widths
measured for the same star by \citet{cayrel2004}.  We find a mean
difference of $-$0.6 m{\AA} with a $\sigma = 2.4$ m{\AA} between the
MIKE and \citet{cayrel2004} measurements for lines included in our
analysis.  We also made a comparison between the equivalent widths
measured from the HRS and MIKE spectra for a representative star, as
we did not have an HRS spectrum for BD~$-$18~5550 because it is not
observable from McDonald Observatory.  We find a mean difference of
2.5 m{\AA} with $\sigma = 3.4$ m{\AA}.  In both comparisons, between
the MIKE spectra measured equivalent widths and those of the
\citet{cayrel2004} study as well as between the MIKE spectra and the
HRS spectra, the measured offset between the measurements shows no
significant disagreement between the techniques.

\subsection{Analysis Techniques}\label{analtech}
The large number of stars in the full $\sim500$ star CASH snapshot
sample calls for automation of the analysis.  Stellar parameters and
elemental abundances from the snapshot HRS spectra were determined
using our newly developed spectroscopic stellar parameter and
abundance analysis pipeline, Cashcode.  The pipeline is written around
the existing platform of the local thermodynamic equilibrium (LTE)
stellar line analysis and spectrum synthesis code MOOG (the latest
version, 2010; \citealt{moog}).  The most recent version of MOOG
accounts for the fact that Rayleigh scattering becomes an
important source of continuum opacity at short wavelengths, blueward
of 4500\,{\AA}.  This is important for our sample stars, as the effect
is more pronounced in cool giants \citep{newmoog}.

We compared the results of four representative stars, two from the
pilot sample and two standard stars, using the newest version of MOOG
and an older version that did not distinctly deal with Rayleigh
scattering in the calculation of the continuum opacities.  We find
that the spectroscopic effective temperatures and microturbulences are
lower, thus the derived [Fe/H] abundances are lower in the version
that deals with Rayleigh scattering by $\sim0.1$-$0.2$ dex.
Generally, the abundance ratios [X/Fe] remain within $\sim0.05$ dex.
These abundances were determined using lines down to
$\sim3750$\,{\AA}; however, for studies with a spectral range that
encompasses shorter wavelengths, this effect may be larger.
Table~\ref{2002comp} shows the stellar parameter comparison and
Table~\ref{2002abund} shows the abundance ratio comparison.

\begin{deluxetable*}{lcccc|cccc}
  \tablecolumns{9}
  \tablewidth{0pc}
  \tablecaption{Stellar Parameter Comparison}\label{2002comp}
  \tablehead{\colhead{}&
\multicolumn{4}{c}{Scattering Treatment Included}&\multicolumn{4}{c}{No Scattering Treatment}\\
\colhead{Star}&\colhead{T$_{eff}$}&\colhead{log g}&\colhead{$\xi$}&\colhead{$\mbox{[Fe/H]}$}&\colhead{T$_{eff}$}&\colhead{log g}&\colhead{$\xi$}&\colhead{$\mbox{[Fe/H]}$}\\
\colhead{}&\colhead{[K]}&\colhead{}&\colhead{[km/s]}&\colhead{}&\colhead{[K]}&\colhead{}&\colhead{[km/s]}&\colhead{}}

\startdata
HD~122563      & 4450 & 0.50 & 2.30 & $-$2.96 & 4475 &  0.40 & 2.55 & $-$2.85\\
CS~22891$-$200 & 4500 & 0.45 & 2.60 & $-$3.93 & 4600 &  0.65 & 3.00 & $-$3.72\\
HE~0015+0048   & 4600 & 0.90 & 1.85 & $-$3.07 & 4675 &  1.05 & 2.15 & $-$3.00\\
HE~0432$-$1000a& 4525 & 0.50 & 2.00 & $-$3.21 & 4600 &  0.65 & 2.60 & $-$3.08\\
\enddata
\label{2002comp}
\end{deluxetable*}

\begin{deluxetable*}{lrr|rr|rr|rr} 
  \tablecolumns{20} 
  \tablewidth{0pc} 
  \tablecaption{Abundance Comparison for Different Treatments of Scattering In MOOG}
  \tablehead{\colhead{}&\multicolumn{2}{c}{CS~22891$-$200}&\multicolumn{2}{c}{HD~122653}&\multicolumn{2}{c}{HE~0015+0048}&\multicolumn{2}{c}{HE~0432$-$1000a} \\
\colhead{Elem}&\colhead{[X/Fe]$_{\rm{scat}}$}&\colhead{[X/Fe]$_{\rm{non}}$}&\colhead{[X/Fe]$_{\rm{scat}}$}&\colhead{[X/Fe]$_{\rm{non}}$}&\colhead{[X/Fe]$_{\rm{scat}}$}&\colhead{[X/Fe]$_{\rm{non}}$}&\colhead{[X/Fe]$_{\rm{scat}}$}&\colhead{[X/Fe]$_{\rm{non}}$}}
\startdata
$\mbox{[Mg/Fe]}$ &0.53   & 0.66    & 0.54  &  0.41 & 0.65  &0.69&0.50  &0.52 \\
$\mbox{[Ca/Fe]}$ &0.68   & 0.72    & 0.39  &  0.36 & 0.44  &0.43&0.39  &0.35 \\
$\mbox{[Cr/Fe]}$ &$-$0.43& $-$0.44 &$-$0.18&$-$0.23&$-$0.17&$-$0.15&$-$0.15&$-$0.18\\
$\mbox{[Ni/Fe]}$ &$-$0.08& $-$0.31 & 0.19  &  0.17 &$-$0.02&0.05 & 0.01 & 0.16\\

\enddata 
\label{2002abund}
\end{deluxetable*} 

\section{Stellar Parameters}\label{stellarparams}
The first step in our abundance analysis is to determine the
atmospheric parameters of each star.  We accomplished this in two
ways: with the Cashcode pipeline and the traditional, manual way.

In order to test the robustness of the snapshot abundances one must
answer two questions: first, for a given set of stellar parameters,
with what precision can abundances be determined from snapshot
spectra?  Secondly, does the pipeline give reasonable stellar
parameters (effective temperature, surface gravity, metallicity, and
microturbulent velocity) for snapshot spectra?  The first question has
been partially answered by the HERES study, which finds that the
uncertainties are $\sim0.25$ dex, but the systematic uncertainties
associated with the stellar parameter and abundance determination
methods and the inherent scatter of our snapshot data set must be
explored.  This is addressed in Section~\ref{syserrors}.  The second
question can be answered by testing the pipeline in detail and
comparing the results from the literature with previously studied
standard stars of comparable stellar parameters and metallicity; this
is discussed in Section~\ref{complit}.  Table~\ref{stellpar} lists our
adopted stellar parameters.

\begin{deluxetable*}{lcccc|cccc} 
\tablecolumns{9} 
\tablewidth{0pc} 
\tablecaption{Stellar Parameters\label{stellpar}}
\tablehead{\colhead{}&\multicolumn{4}{c}{MIKE}&\multicolumn{4}{c}{HRS}\\
\cline{2-5} \cline{6-9}\\
\colhead{Star}&\colhead{T$_{eff}$}&\colhead{log g}&\colhead{$\xi$} &\colhead{[Fe/H]}&\colhead{T$_{eff}$}&\colhead{log g}&\colhead{$\xi$} &\colhead{[Fe/H]}  \\
\colhead{}  & \colhead{[K]} &\colhead{}& \colhead{[km/s]}&\colhead{} & \colhead{[K]} &\colhead{}& \colhead{[km/s]}&\colhead{} }
\startdata
HE~0013$-$0257  & 4500 &  0.50  &2.10  & $-$3.82  &  4710  &  1.35  & 2.25  & $-$3.40  \\
HE~0013$-$0522  & 4900 &  1.70  &1.80  & $-$3.24  &  5120  &  2.30  & 2.00  & $-$2.90  \\
HE~0015$+$0048  & 4600 &  0.90  &1.85  & $-$3.07  &  4630  &  1.10  & 1.90  & $-$3.00  \\
HE~0302$-$3417a & 4400 &  0.20  &2.00  & $-$3.70  & \nodata& \nodata&\nodata&\nodata   \\
HE~0324$+$0152a & 4775 &  1.20  &1.80  & $-$3.32  &  4800  &  1.65  &  2.00 & $-$3.05  \\
HE~0420$+$0123a  & 4800 &  1.45  &1.50  & $-$3.03  &  4800  &  1.45  &  1.80 & $-$3.00  \\
HE~0432$-$1005a & 4525 &  0.50  &2.00  & $-$3.21  &  4540  &  0.65  &  2.00 & $-$3.10  \\
HE~1116$-$0634  & 4400 &  0.10  &2.40  & $-$3.73  &  4650  &  1.00  &  3.80 & $-$3.35\tablenotemark{a}\\
HE~1311$-$0131  & 4825 &  1.50  &1.95  & $-$3.15  &  4820  &  1.50  &  2.00 & $-$2.85  \\
HE~1317$-$0407  & 4525 &  0.30  &2.15  & $-$3.10  &  4600  &  0.25  &  3.20 & $-$3.10\tablenotemark{a}\\
HE~2123$-$0329  & 4725 &  1.15  &1.80  & $-$3.22  &  4700  &  1.40  &  2.00 & $-$3.05  \\
HE~2138$-$0314  & 5015 &  1.90  &1.75  & $-$3.29  &  4940  &  2.30  &  2.00 & $-$3.20  \\
HE~2148$-$1105a & 4400 &  0.20  &2.65  & $-$2.98  &  4450  &  0.20  &  3.20 & $-$3.05\tablenotemark{a}\\
HE~2238$-$0131  & 4350 &  0.15  &2.45  & $-$3.00  &  4300  &  0.30  &  2.15 & $-$3.05  \\
HE~2253$-$0849  & 4425 &  0.20  &2.65  & $-$2.88  &  4420  &  0.15  &  2.30 & $-$2.85  \\
HE~2302$-$2154a & 4675 &  0.90  &2.00  & $-$3.90  & \nodata& \nodata&\nodata&\nodata   \\
CS~22891$-$200  & 4500 &  0.45  &2.60  & $-$3.92  & \nodata& \nodata&\nodata&\nodata   \\
HD~122563       & 4450 &  0.50  &2.30  & $-$2.96  &  4400  &  0.30  & 2.25  & $-$2.95  \\
BD~$-$18~5550   & 4600 &  0.80  &1.70  & $-$3.20  & \nodata& \nodata&\nodata&\nodata   \\
CD~$-$38~245     & 4650 &  0.95  &2.15  & $-$4.00  & \nodata& \nodata&\nodata&\nodata   \\
CS~22873$-$166  & 4375 &  0.20  &2.80  & $-$3.14  & \nodata& \nodata&\nodata&\nodata   \\
\enddata 
\tablenotetext{a}{Sky contamination in this spectrum}
\end{deluxetable*} 

\subsection{HRS Snapshot Data}\label{sec:hrs_snaps}
The effective temperature of a star is spectroscopically determined by
minimizing the trend of the relation between the abundance and
excitation potential of the lines from which the abundance is derived.
The microturbulent velocity is determined by doing the same for the
abundance and reduced equivalent width.  The surface gravity is
determined from the balance of two ionization stages of the same
element (e.g.,~Fe I and Fe II).  The HRS snapshot spectra have few
detectable Fe II lines, thus we use Ti I and Ti II lines in addition
to Fe I and Fe II lines in the pipeline to more robustly determine the
stellar parameters, in particular the gravity, from these spectra,
with Fe weighted twice compared to Ti.  Often, only the ionization
balance of Fe is considered.  The metallicity used in the model
atmosphere, in this case [M/H], is an average of the abundances from
individual lines of Fe I, Fe II, Ti I, and Ti II.

The pipeline works iteratively.  The first process is to determine the
stellar parameters from the equivalent width measurements of Ti I, Ti
II, Fe I, and Fe II.  An approximate initial guess to the effective
temperature, surface gravity, metallicity, and microturbulent velocity
are input as well as constraints on the parameters over which the code
iterates.  Generally for the HRS snapshot data, we require that the
trend between abundance and excitation potential is $<|0.03|$ dex/eV,
the reduced equivalent width and abundance trend is $<|0.15|$
dex/log(m{\AA}), the surface gravity criterion $\Delta_{ion} < |0.10|$
dex, and the difference between the metallicity of the model
atmosphere and the calculated metallicity is $<|0.10|$ dex.  As a
result of the ionization balance constraints, the abundance difference
between Ti I and Ti II is allowed to be no greater than 0.3 dex if the
Fe I abundance equals the Fe II abundance or the abundance difference
between Fe I and Fe II is allowed to be no greater than 0.2 dex if the
Ti I abundance equals the Ti II abundance.  This information is used
to construct an initial Kurucz stellar atmosphere with
$\alpha$-enhancement \citep{CastelliKurucz} to begin the stellar
parameter determination, in which the pipeline iterates until the
various constraints that determine each stellar parameter fall within
the user-defined thresholds specified in the beginning.

We used the pipeline to derive abundances for each line from its
equivalent width.  We used synthetic spectra to determine abundances
for particular lines (e.g.,~Ba II $\lambda4554$) for various reasons,
including hyperfine structure and blending with other features.  For
these lines, we used an equivalent width-derived abundance guess as an
initial input for the spectral syntheses.

\subsection{High-Resolution MIKE Spectra}\label{hiresmike}
The high-resolution MIKE data were analyzed manually.  The large
wavelength coverage allowed us to perform a more in-depth analysis with
smaller uncertainties for a later comparison to the pipeline analysis
of the corresponding snapshot spectra.  Spectroscopic stellar
parameters for all but two of the MIKE spectra were determined from
equivalent width measurements of Fe I and Fe II lines.  The resonance
lines of Fe I were excluded in this analysis, as they often are near
the flat portion of the curve of growth.  In essence, all of the steps
in the Cashcode pipeline were performed, but each step was executed
manually.  This allowed us to compare the pipeline
results with those of a manual analysis to confirm that the pipeline
reproduced the results derived from the high-resolution data.

It is also possible to determine effective temperatures
photometrically, using calibrations between colors and temperature for
given color and metallicity ranges.  We have chosen to adopt
spectroscopic temperatures for these stars, as well as for the entire CASH
sample, in order to present a homogeneous set of atmospheric
parameters and, therefore, the resulting chemical abundances.
However, we determined photometric temperatures to check the accuracy
and systematic uncertainties of our method.  Accurate long baseline
colors (i.e., V$-$K) for the entire $\sim500$ star sample do not
exist; however, we used 2MASS photometry for the pilot sample and
standard stars in order to determine photometric temperatures in order
to compare them to the adopted spectroscopic temperatures.

\begin{figure}[]
\includegraphics[width=0.5\textwidth]{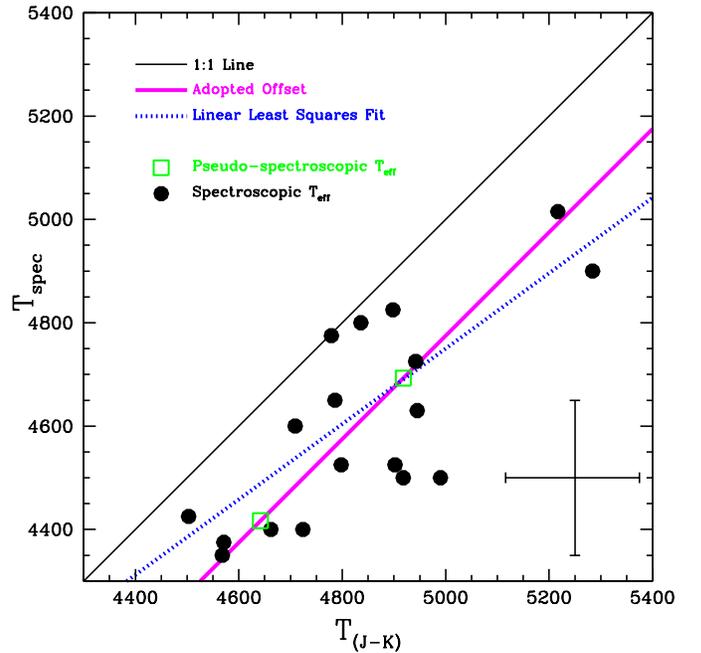}
\caption{Comparison of the spectroscopic temperatures to those derived
  from (J$-$K) 2MASS photometry using the \citet{Alonso_giants}
  calibration.  The thin black line is the 1:1 comparison, the dotted blue
  line is the linear least squares fit to the data, and the thick pink
  line shows the adopted offset applied to HE~1116$-$0634 and
  HE~2302$-$0317a, which are both represented by the green square
  points.}
\label{phot}
\end{figure}

We determined reddening corrections for our stars using the
\citet{schlegel} dust maps.  We dereddened the J$-$K 2MASS color and
then, according to equations 1a-1c in \citet{ramirezmelendez2004},
transformed J and\,K 2MASS photometry into the TCS system in order to
use the \citet{Alonso_giants} calibration to determine photometric
temperatures.  We determined the formal linear relation between the
(J$-$K) photometric temperatures from the sample stars and the
spectroscopic temperatures.  Within the uncertainty of the fit, there
exists an offset between the spectroscopic and photometric
temperatures.  Thus, we adopted the mean difference between the
spectroscopic and photometric temperatures to be T$_{spec} =
$T$_{(J-K)}-225$\,K.  The uncertainties associated with the
spectroscopic and photometric temperatures are 160 and 140\,K,
respectively.  See the next subsection for further details.
Figure~\ref{phot} shows the spectroscopic temperature plotted against
the (J$-$K) photometric temperature for the sample stars.  Plotted are
lines that show a 1:1 agreement, the adopted offset, and the least
squares fit.

Two stars in the sample, HE~1116$-$0634 and HE~2302$-$2154a, did not
have convergent spectroscopic stellar parameter solutions irrespective
of the analysis method.  The Fe I and Fe II abundances were not in
agreement (i.e., the surface gravity criterion was not met) when the
temperature criterion was met. This is likely because these stars are
near the edge of the stellar atmosphere grid in terms of metallicity,
temperature, and gravity.  For these stars, we adopted a
``pseudo-spectroscopic'' effective temperatures of 4400 and 4675\,K
for HE~1116$-$0634 and HE~2302$-$2154a, respectively.  Those were
obtained by applying the previously determined offset between the
spectroscopic and photometric temperatures of 225\,K to their
photometric temperatures.  The pseudo-spectroscopic temperatures were
used to determine the remaining parameters, surface gravity,
microturbulence, and metallicity spectroscopically, ignoring the
nonzero slope of $-0.065$ and $-0.090$ dex/eV for HE~1116$-$0634 and
HE~2302$-$2154a, respectively, of the excitation potential. Typically,
this slope is $<0.02$ dex/eV.

\begin{figure}[]
\includegraphics[width=0.5\textwidth]{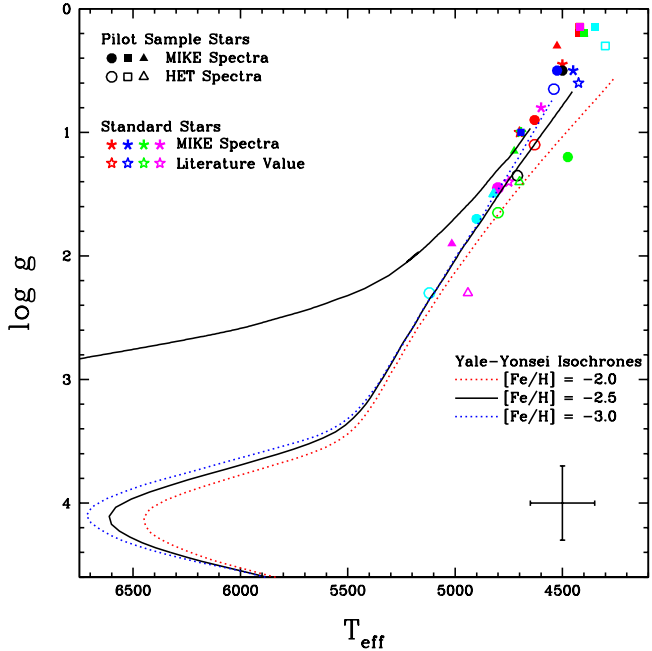}
\caption{HR diagram of the pilot sample and standard stars.  For the
pilot sample, open symbols represent the stellar parameters derived
from the HRS spectra and the solid symbols represent stellar
parameters derived from the MIKE spectra.  For the standard stars,
open stars represent stellar parameters from the literature and
asterisks represent stellar parameters derived from the MIKE spectra.
Overplotted are the Yale-Yonsei isochrones \citep{green,Y2_iso} for 12
Gyr, at [Fe/H] = $-$2.0 (red line), $-$2.5 (black line), and $-$3.0
(blue line) as well as a horizontal-branch mass track from
\citet{HBtrack}.\label{hrd}}
\end{figure}

Figure~\ref{hrd} shows the derived effective temperatures and surface
gravities for all stars in the sample plotted against 12 Gyr
Yale-Yonsei isochrones \citep{green, Y2_iso} for [Fe/H] = $-2.0$,
$-2.5$, and $-3.0$, as well as a \citet{HBtrack} horizontal-branch
mass track.  For the standard stars, we show the stellar parameters
derived from the MIKE spectra, as well as literature values taken from
\citet{cayrel2004} for BD~$-$18~5550, CD~$-$38~245, and
CS~22873$-$166, \citet{McWilliametal} for CS~22891$-$200, and
\citet{fulbright} for HD~122653.  Due to the low-metallicity of
CD~$-$38~245, the corresponding MIKE spectrum shows very few
absorption lines, thus we included this star only in the calibration
of the stellar parameter offset.

\subsection{Uncertainties}
Each star has two or three measurements of temperature: spectroscopic
temperatures derived from the MIKE and HRS spectra (when available)
and a photometric temperature based on JHK 2MASS colors using the
\citet{Alonso_giants} calibration.  

We determined random uncertainties in the spectroscopic temperatures
based upon the uncertainty in the slope determined by the Fe I line
abundances.  For a representative star, we varied the temperature
until the resultant Fe abundance was one standard deviation away from
the original derived abundance.  We found this value to be $\sim
125$\,K; we adopt this as our random uncertainty.  We determined
random uncertainties in the photometric temperatures based upon the
uncertainties given for the 2MASS colors.  We found this uncertainty
be $\sim140$\,K.

We also compared the derived spectroscopic and photometric
temperatures.  We remind the reader that the offset between the two
temperature scales is $\sim225$\,K, where $\Delta T_{eff}$ =
$T_{eff}^{Phot}-T_{eff}^{Spec}$.  Given this offset, we derive a
systematic uncertainty for our spectroscopic temperatures of $\sim160$
\,K.  This is of the same order as the random uncertainties.

We obtained the random uncertainty in the surface gravity by allowing
the Fe I and Fe II values to vary until they no longer agree within
the uncertainty of Fe I, which is $\sim0.25$ dex.  Since all the pilot
sample stars are on the giant branch, uncertainties in effective
temperature at the $\sim150$\,K level lead to changes in the surface
gravities of $\sim 0.5$ dex.  We conservatively adopt this as our
$\sigma_{log g}$ uncertainty.

We calculated the standard error of the mean Fe I abundance, obtained
from individual Fe I line abundances and found it to be $\sim0.01$
dex. This is rather small, as the calculation does not take continuum
placement uncertainties into account.  With that in mind, we adopt the
scatter of the individual Fe line abundances as a more conservative
random [Fe/H] uncertainty, which is $\sim0.12$ dex, although it varies
slightly by star.

In order to ascertain the uncertainty in the microturbulence, we
determined the maximum change to the microturbulence values which would
still yield the same [Fe/H] value within the uncertainties. This leads
to $\sigma_{\xi} \sim0.3$ km/s.

\section{Robustness of the Stellar Parameter and Abundance Pipeline}\label{robustsp}
\subsection{Stellar Parameters}
One of the main purposes of this paper is to test the stellar
parameter and abundance pipeline that will be used for the larger CASH
sample. The first test is to compare the manually derived stellar
parameter results with those determined with the Cashcode pipeline.
This can be done in three ways: i) comparing the stellar parameters and
abundances of a manual analysis of a given spectrum with each Cashcode
result; ii) comparing the results derived from snapshot spectra taken of
standard stars with those of well determined literature values of the
stellar parameters; and iii) comparing the stellar parameter results of a
snapshot spectrum with those of a corresponding high-resolution
spectrum.

We first tested the accuracy of the pipeline and the precision of our
iteration criteria by comparing the results derived from a
representative manual analysis of a high-resolution MIKE spectrum to
those derived using the pipeline.  We find $\Delta$T$_{eff}$=10\,K,
$\Delta$log g = 0.0, $\Delta\xi$ = 0.05, and $\Delta$[Fe/H]= 0.02 dex.
We find that the [X/Fe] values (where X is a given element) agree to
within $\sim0.05$ dex.

By comparing the Cashcode results of standard stars to the literature
values, we can test how accurately the snapshot data are able to
reproduce the results of independent studies derived from
high-resolution, high-S/N data using traditional manual methods.  We
evaluated an absolute consistency between the Cashcode results and the
literature.  See Subsection~\ref{complit} for details on the abundance
differences.  A caveat to this comparison is that standard stars are
usually bright targets, such that their snapshot spectra have much
higher S/N than the median value for a snapshot spectrum.  Generally,
Cashcode produces abundance results with smaller uncertainties for
higher-S/N, higher-resolution data because in these cases the line
abundance scatter is decreased, therefore the user-defined parameter
fitting criteria can be tightened.

However, the purpose of this study is to test the pipeline for the
median S/N snapshot star.  Unfortunately, we did not observe a low S/N
snapshot spectrum of the standard stars, which precludes the best
possible comparison.  To resolve the issue, we instead turn to
high-resolution data for testing the pipeline for a median S/N
snapshot data.  We compared the HRS pipeline-derived results to those
of the manual, high-resolution analysis for all stars.  We find
agreement in the stellar parameters to be within $\Delta T_{eff}\pm
55$\,K, $\Delta$ log g$\pm$ 0.3 dex, $\Delta\mbox{[Fe/H]}\pm 0.15$ dex,
and $\Delta\xi\pm 0.21$ dex.

\subsection{Chemical Abundances}
For each star with an HRS spectrum, we compared the derived abundances
with the high-resolution MIKE spectrum.  For the 12 stars in common, we
found the offset between the two to have a standard error, or standard
deviation of the mean, of 0.07 dex.  The average $\Delta
{\mbox{\rm{[X/Fe]}}}$ over all elements was 0.09 dex, thus the
abundances derived from the HRS data with Cashcode can reproduce a
manual analysis to within 1.5$\sigma$.  Over all the elements, there
is no statistically significant difference between the HRS/Cashcode
and the MIKE/manual analysis.  For individual elements, some
discrepancies do exist.  We found that the largest discrepancies
between the HRS and MIKE spectra arose for those elements whose lines
occurred in regions of low S/N (e.g.,~the $\Delta
{\mbox{\rm{[X/Fe]}}}$ for Sr II is larger than that of Ba II, as the
only Sr II line in the HRS occurs at 4215\,{\AA}, while there are Ba
II lines at longer wavelengths).

\begin{figure*}[]
\includegraphics[width=1.0\textwidth]{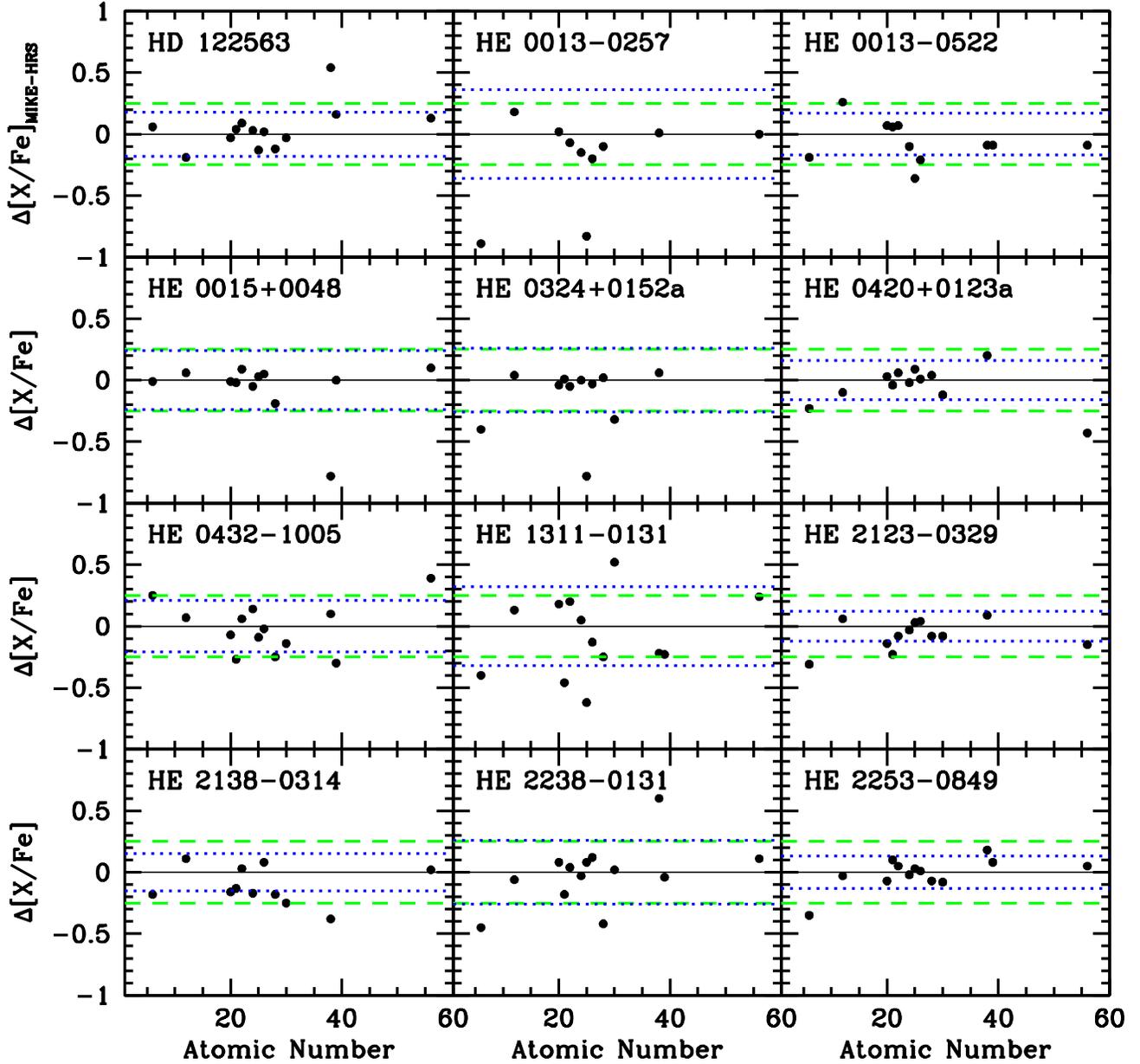}
\caption{Abundance differences, $\Delta$[X/Fe], defined as
  [X/Fe]$_{\rm{MIKE}}$ - [X/Fe]$_{\rm{HRS}}$, shown as a function of
  atomic number for each star that was observed with both HRS and
  MIKE.  The black solid line represents zero offset, the green dashed
  line represents the 0.25 dex random error derived in the HERES
  study, and for comparison, the blue dotted line represents the
  calculated spread for each star.
\label{mikecashcomp}}
\end{figure*}

\begin{figure*}[]
\includegraphics[width=1.0\textwidth]{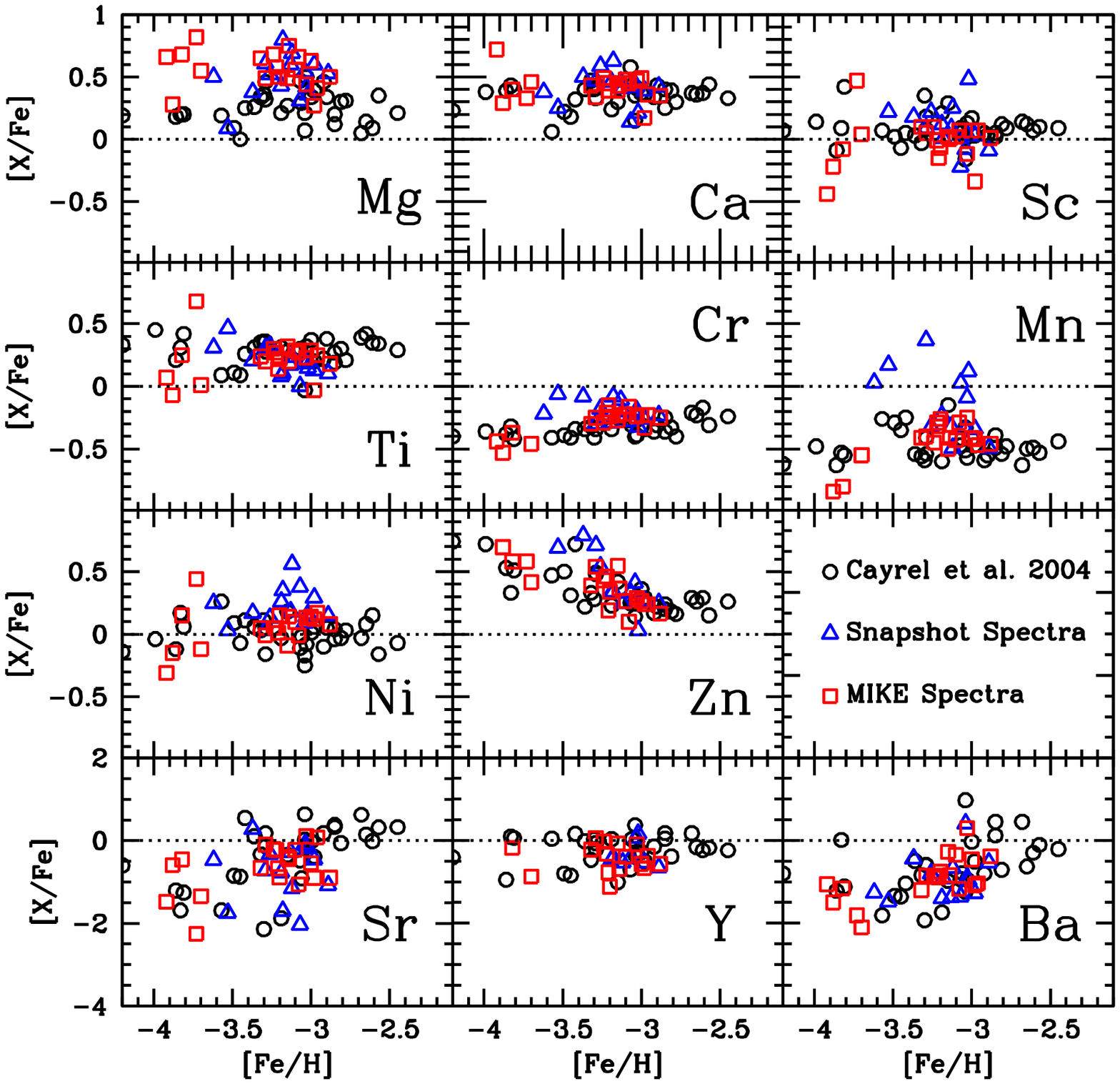}
\caption{[X/Fe] abundance ratios vs. [Fe/H] for each of the elements
  measured using Cashcode in the HRS snapshot spectra (blue triangles)
  compared with the MIKE abundances (red squares) and the
  \citet{cayrel2004} abundances (black circles). The black dotted line
  represents the solar abundance ratio.  The HRS Ni abundances shown
  are preliminary.  \label{cashrel}}
\end{figure*}

For a given element, there is a discrepancy in the measured abundances
derived from HRS and MIKE spectra. These range from 0.08 dex, in the case
of Ti, which has 9 lines across the HRS spectrum, to 0.36 dex, in the
case of Sr, which has one feature at 4215\,{\AA} in the HRS spectrum.
This value is dependent upon the location in wavelength and number of
lines per element; bluer features from lower S/N spectral regions have
larger discrepancies.  Figure~\ref{mikecashcomp} illustrates the
$\Delta {\mbox{\rm{[X/Fe]}}}$ for each element that was measured in
both the HRS and MIKE spectrum of a particular star.  We also compared
the full set of [X/Fe] CASH and MIKE derived abundances to those of
the \citet{cayrel2004} study.  Figure~\ref{cashrel} shows the
comparison between the three data sets.  Manganese abundances are
sometimes largely discrepant; however, this is likely because there
are only four Mn lines available to detection in the HRS spectra and
often only one of these lines was detected.  It has been previously
noted \citep{cayrel2004, roedererlinelist} that the line selection for Mn is
critical in understanding the derived abundances and the
\citet{cayrel2004} study included additional lines not available in
the HRS spectra.

\section{Abundance Analysis}\label{abundanal}
We present chemical abundances and upper limits for 13 elements
derived from the HRS spectra and 18 elements from the MIKE spectra in
Table~\ref{sampletbl}.  We now describe the details of our abundance
analysis.  For each element, we discuss the method of abundance
determination, the relevant spectral features, the number of stars in
which this element was measured, and any differences in the analysis
between the high-resolution and snapshot spectra.

\begin{deluxetable}{lrrrrr}
  \tablecolumns{20}
  \tablewidth{0pc}
  \tablecaption{Abundances\label{sampletbl}}
  \tablehead{\colhead{Elem}&\colhead{log$\epsilon$(X)}&\colhead{[X/Fe]}&\colhead{$\sigma$}&\colhead{n}&\colhead{[X/Fe]}\\
\colhead{}&\colhead{dex}&\colhead{MIKE}&\colhead{dex}&\colhead{}&\colhead{HRS}\\
\\
\multicolumn{6}{c}{HE~0015-0048}}
\startdata
Mg I   & 5.18  & 0.66   &0.18  &6  & 0.62\\
Ca I   & 3.71  & 0.45   &0.14  &15 & 0.44\\
Sc II  & 0.14  & 0.07   &0.14  &8  & 0.06\\
Ti I   & 1.98  & 0.11   &0.18  &18 & 0.11\\
Ti II  & 2.14  & 0.27   &0.15  &39 & 0.06\\
Cr I   & 2.40  & $-0.16$&0.20  &14 & $-$0.14\\
Mn I   & 2.06  & $-0.29$&0.14  &4  & $-$0.28\\
Co I   & 1.97  &0.06    &0.14  &6  & \nodata\\
Ni I   & 3.13  & $-0.01$&0.16  &8  & \nodata\\
Zn I   & 1.58  & 0.10   & 0.14 & 1 &0.68 \\
Sr II  &$-$1.06& $-$0.87& 0.14 & 2 & $-0.39$\\
Ba II  &$-$2.07& $-$1.17& 0.14 & 3 &$-1.30$\\
\enddata
\tablecomments{Table~\ref{sampletbl} is published in its entirety in the electronic edition.  A
portion is shown here for guidance regarding its content.}
\end{deluxetable}

\begin{figure*}[]
\includegraphics[width=1.0\textwidth]{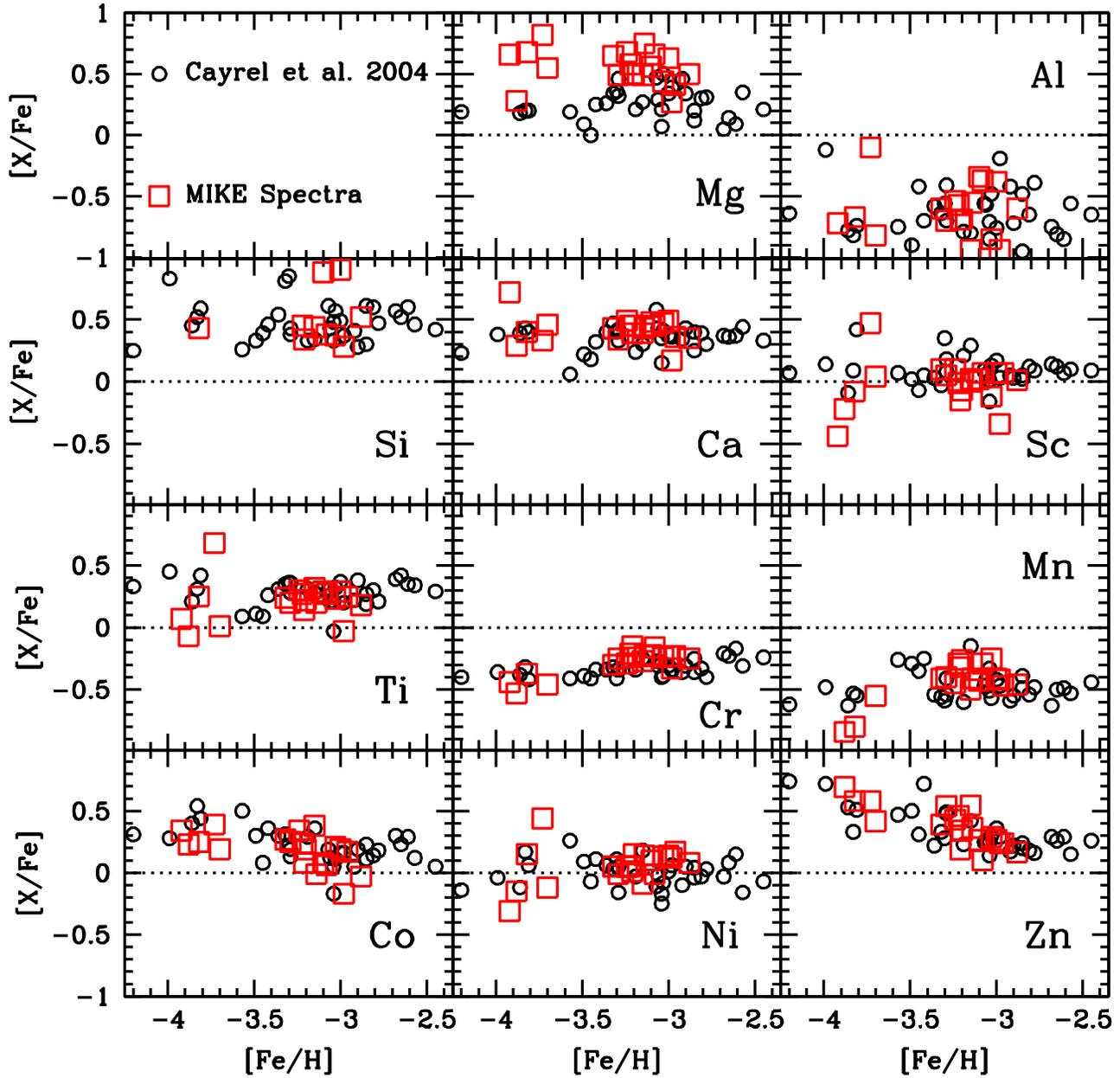}
\caption{[X/Fe] abundance ratios vs. [Fe/H] for each of the
  elements up to Zn manually measured in the MIKE spectra (red
  squares) compared with the \citet{cayrel2004} abundances (black
  squares). The black dotted line represents the solar abundance
  ratio. \label{mikerel}}
\end{figure*}

\begin{figure*}[]
\includegraphics[width=1.0\textwidth]{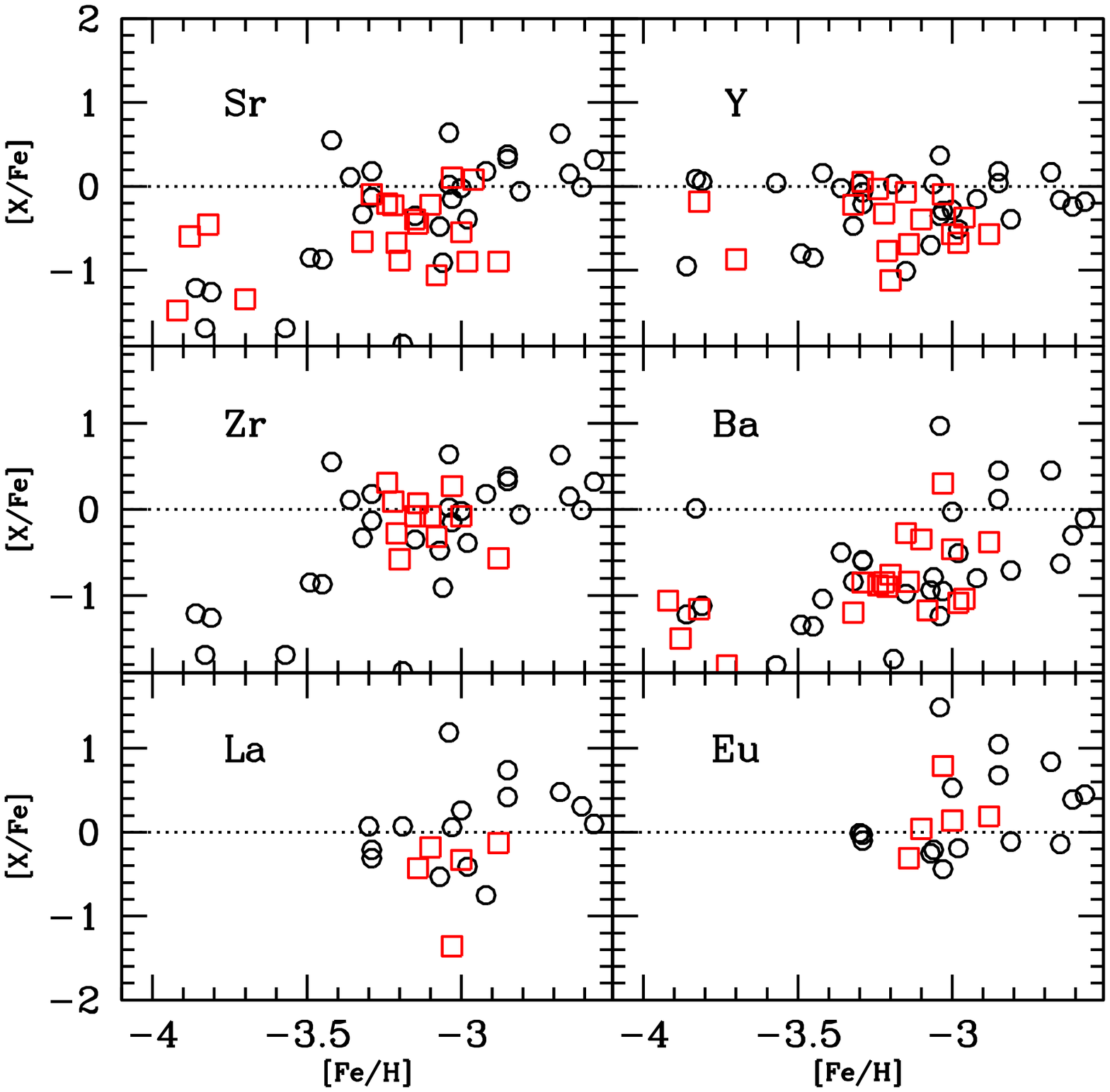}
\caption{[X/Fe] abundance ratios vs. [Fe/H] for six neutron-capture
  elements measured in the MIKE spectra (red squares) compared with
  the \citet{cayrel2004} abundances (black circles). The black dotted
  line represents the solar abundance ratio.\label{mikencap}}
\end{figure*}

\subsection{Light Elements: Li, C, Al, Si}
Lithium abundances were determined from both the HRS and MIKE spectra
through synthesis of the Li I doublet at 6707\,{\AA}.  Both spectra
cover this feature.  Lithium is most easily detected in main-sequence
stars, but as a star ascends the RGB, its Li abundance
becomes depleted.  Thus, we only expect to detect the Li feature in
our warmest giants.  We detected the doublet in 5 of 12 stars stars
with HRS spectra and 7 of 20 stars MIKE spectra.

Carbon abundances were determined from the CH G-band feature at
4313\,{\AA}.  The S/N near the G-band head region in the HRS snapshot
spectra is low, such that we have to perform a manual synthesis of the
G-band head in order to determine the C abundances.  At low S/N,
Cashcode does not yield reliable results mostly due to uncertainties
in continuum placement.  For synthesis we assumed an [O/Fe] abundance
ratio of 0.0.  When the [O/Fe] ratio reaches $\sim1$, the C abundances
determined from molecular features is affected, though none of our
stars indicate such a high O abundance.  We detected this feature in
all stars in both sets of spectra.  The two available O indicators in
our spectra are the forbidden line at 6300\,{\AA} and the O triplet.
At the metallicity of our sample, we are unable to detect the
forbidden line except for cases of extreme O enhancement.  The triplet
lines are known to have NLTE effects and thus we did not measure an O
abundance.

Sodium abundances were not included in this analysis as we have not
implemented a routine in the pipeline to discern the stellar
absorption lines from the interstellar emission.

Aluminum abundances were determined using equivalent width
measurements of the $\lambda$3961 line of the Al I resonance doublet.
We opted to use only the $\lambda$3961 line for our abundance analysis
because the other line, $\lambda$3944, is blended with CH.  Neither
feature falls within the HRS snapshot wavelength regime.  We measured
this line in 19 of the MIKE spectra.  We did not measure the Al
abundance in HD~122563 due to the low S/N of that spectrum.

Two Si lines are detectable in metal-poor stars: $\lambda$3906 and
$\lambda$4103.  The $\lambda$3906 line is heavily blended with CH in a
low S/N region of the spectrum; however, we prefer to use only unblended lines
for our analysis. The $\lambda$4103 line falls within the wing of the
H$\delta$ line, thus we adopt a local continuum in the wing of the
H$\delta$ line.  We present spectral synthesis derived abundances from
the MIKE spectra using the $\lambda$4103 line. Neither feature is
available in the snapshot spectra due to limited wavelength coverage.
We measured this line in 19 stars with MIKE spectra.

\subsection{$\alpha$-Elements: Mg, Ca, Ti}
Abundances for Mg, Ca, and Ti were determined for all stars in the
sample from both the HRS and MIKE spectra from equivalent width
analysis.  Only 4 unblended Mg I lines are available in the HRS
snapshot spectra: $\lambda$4703, $\lambda$5173, $\lambda$5184, and
$\lambda$5528.  The Mg b lines are often near the flat part of the
curve of growth, but their abundances generally agree with those of
the abundances of the other two lines.  Thus, they were included in
the abundance determination.  There are $\sim12$ detectable lines in
the MIKE spectra, although in these spectra we exclude lines on the
flat part of the curve of growth.  This is determined on a
line-by-line basis for each star.

There are four available Ca I lines in the snapshot spectra:
$\lambda$5589, $\lambda$6103, $\lambda$6122,
$\lambda$6162.  Of the 4 lines, the latter 3 are sometimes
blended with telluric features.  This was assessed on a line-by-line
basis in each star.  There are $\sim30$ Ca I lines in the MIKE
spectra.  Though the Ca I $\lambda$4226 line is available in the HRS spectra,
we did not include it in our analysis because of the low S/N of the
feature.  Additionally, this line is often excluded from the MIKE
analysis because it is in the flat part of the curve of
growth.  

Titanium abundances were determined from 6 Ti I and 9 Ti II lines in
the snapshot spectra and $\sim30$ Ti I lines and $\sim60$ Ti II in the
high-resolution spectra.

\subsection{Fe-Peak Elements: Sc, Cr, Mn, Co, Ni, Zn}
All abundances were determined from equivalent width measurements
except for Zn because it has only two weak lines.  There are 4
detectable Sc II lines in the snapshot data: $\lambda$5031,
$\lambda$5239, $\lambda$5526, and $\lambda$5657.  There are $\sim15$
Sc II lines detectable in the MIKE spectra.  We detected Sc lines in 20
MIKE spectra and 11 HRS spectra.

We measured five Cr I lines in the snapshot spectra: $\lambda$4616,
$\lambda$4646, $\lambda$5206, $\lambda$5346, and $\lambda$5410.  We
measured $\sim20$ Cr I lines and 4 Cr II lines in the MIKE spectra:
$\lambda$3409, $\lambda$4559, $\lambda$4588, and $\lambda$4592.  There
is a difference between the abundances derived between Cr I and Cr II,
with Cr II abundances typically$\sim+0.35$ dex larger than Cr I.  We detected
the Cr I lines in 19 MIKE spectra and 12 HRS spectra.  In all plots,
we adopt the $\mbox{[Cr I/Fe]}$ values as [Cr/Fe].

We measured 4 Mn I lines in the HRS spectra:
$\lambda$4754, $\lambda$4762, $\lambda$4783, and 
$\lambda$4823.  We measured 8 Mn I lines in the
MIKE spectra.  We detected Mn I lines in 19 MIKE spectra and 11 HRS
spectra.

In the MIKE data we measured $\sim30$ Ni I lines, but only the $\lambda$5477
and $\lambda$6177 Ni I lines are available in the HRS spectra.
Abundances derived from the $\lambda$5477 line sometimes are
erroneously large by $\sim0.2$ dex.  We will develop a calibration for
the large CASH sample, but here we only report Ni abundances derived
from the MIKE spectra for all 20 stars.

We measured 11 Co I lines in the MIKE spectra that fall between 3502\,{\AA} and
4122\,{\AA}.  These lines are not covered by the snapshot spectra.  We
detected Co I lines in all 20 MIKE spectra.

Zinc abundances were determined via spectral synthesis of the
$\lambda$4722 and $\lambda$4810 Zn I lines in both the HRS and MIKE
spectra.  We detected Zn I lines in 19 MIKE spectra and 9 HRS
spectra.

\subsection{Neutron-Capture Elements}
We measured abundances for six neutron-capture elements via spectral
synthesis.  The Sr II $\lambda$4215 and $\lambda$4077 lines are
covered by the MIKE spectra, while only the $\lambda$4215 line is
available the HRS spectra.  We detected Sr II lines in all stars in
both the HRS and MIKE spectra.

We synthesized the $\lambda$4883 and $\lambda$5087 Y II lines in the
HRS snapshot spectra and the $\lambda$3949 and $\lambda$4883 lines in
the MIKE spectra.  The $\lambda$3949 line was chosen because it is a
prominent line and can still be detected when a giant star has
sub-solar neutron-capture abundances.  We excluded the $\lambda$5087
line because it falls in a region in the MIKE spectra with low S/N, as
it is at the edge of both the blue and red edges of the CCD.  We
detected Y II lines in 16 MIKE spectra and 6 HRS spectra.  We report
upper limits from the MIKE spectra for 5 stars.

We synthesized the Zr II $\lambda$4209 line to determine the abundance
for the MIKE spectra.  Even though this line is available in the HRS
snapshot spectra, the S/N in the blue region of all spectra is too low
for such a weak feature.  We detected the line in 12 MIKE spectra. We
report upper limits from the MIKE spectra for 9 stars.

For the HRS spectra we used 3 Ba II lines: $\lambda$4554,
$\lambda$5854, and $\lambda$6142; however often only the $\lambda$4554
line can be measured given the low S/N of the HRS data as well as
the low metallicity of the sample.  The same lines were measured in the
MIKE spectra, with the addition of the $\lambda$6496 line.  Generally,
these weaker lines are preferable because the $\lambda$4554 line is
usually on the damping part of the curve of growth, especially in
neutron-capture enhanced stars.  We measured Ba in 20 MIKE spectra and
11 HRS spectra.

We synthesized 2 La II lines at 4086 and 4123\,{\AA} in the MIKE spectra.
The HRS spectra do not cover these wavelengths.  

We synthesized two Eu II lines at 4129 and 4205\,{\AA} in the MIKE
spectra.  The HRS spectra do not cover these wavelengths.  We detected
La II and Eu II features in the 5 same stars in the MIKE spectra.  We
report upper limits from the MIKE spectra for 15 stars for both Eu and
La.

\subsection{Non-LTE Effects}\label{nlte}
Chemical abundances are generally derived under the assumption of one
dimensional (1D) model atmospheres in LTE, but non-LTE effects may
alter our derived values. The non-LTE effects in the elements Mg, Sr,
and Ba have been studied in the stars of the \citet{cayrel2004} sample
by \citet{MgKNLTE}, \citet{SrNLTE}, and \citet{BaNLTE}, respectively.
Based upon their reanalysis of the \citet{cayrel2004} sample, the
non-LTE corrections would be $\sim0.15$ for Mg and Ba in our
sample. In the case of Sr, the non-LTE abundances vary, and can be larger and
smaller than the LTE abundances even for stars of similar temperature
and gravity, so it is difficult to say what this effect would be in
our stars.  See also \citet{asplund_araa} for a comprehensive review
of non-LTE effects on stellar abundances for a range of elements.
Such effects are very sensitive to stellar parameters and individual
lines from which the abundances are derived.  In the absence of ``full
grid'' non-LTE correction calculations, we are not able to apply such
corrections to our sample; however, we remind the reader that these
corrections have implications for interpretations of Galactic chemical
evolution models and should be taken into account for such
investigations.  To illustrate the magnitude of non-LTE effects, we
discuss Al and Mn here, as both have some of the most severe non-LTE
corrections.

\citet{al_nlte} present a non-LTE-corrected [Al/H] abundance analysis
for four sets of stars for the $\lambda$3961, $\lambda$6696,
$\lambda$6698, $\lambda$8772, $\lambda$8773 lines. Only the first line
is detected in stars with [Fe/H] $<-2.0$, and thus applicable to this
study.  The $\lambda$3944 line is also detectable in metal-poor stars,
but it is blended with CH features.  They calculated non-LTE
abundances for 6500\,K and 5200\,K main sequence stars, a 5780\,K solar
analog, and a 5500\,K turn-off star, with [Fe/H] varying from $-3.00$
to 0.00 dex.  For a given set of atmospheric parameters, the
[Al/H]$_{\rm{NLTE}}$ correction increases with decreasing [Fe/H]. The
most evolved star (T$_{eff}$ = 5500\,K , log g = 3.5) in their
analysis has a non-LTE correction of 0.65 dex.  All the stars of the
pilot sample are on the giant branch and generally are more metal poor
than the \citet{al_nlte} models, which indicates that our stars would
have a larger non-LTE correction for the $\lambda$3961 line, though
they would likely all have a similarly large correction.

\citet{mn_nlte} note that Mn has strong NLTE effects, which increase
with decreasing metallicity.  These effects have been shown to be stronger
in the $\lambda4030$ resonance triplet, which we do not include in our
analysis.  The most evolved, metal-poor model analyzed in
\citet{mn_nlte} (T$_{eff}$=5000\,K, log g=4, [Fe/H]=$-3$) has an
average [Mn/Fe] NLTE correction of 0.42 dex for the lines that we
include in our Mn linelist ($\lambda$4030, $\lambda$4033,
$\lambda$4034, $\lambda$4041, $\lambda$4754, $\lambda$4783, and
$\lambda$4823). They also included HD~122563 in their sample of stars
and calculated an NLTE correction of $+0.44$ dex to the [Mn/Fe] ratio.
Again, we do not adopt an NLTE correction for any of our sample,
though corrections of this magnitude would indicate that our average
[Mn/Fe] trend is elevated to a slightly sub-solar level.  The
aforementioned elements have large non-LTE corrections, though other
elements do not have corrections of this magnitude, thus their
abundances can be used for interpretation in a straightforward manner.

\subsection{Uncertainties}\label{syserrors}
To determine the random uncertainty of our abundances, we calculated
the scatter of the individual line abundances for each ionization
state of each element measured.  For any abundance determined from
equivalent width measurements of less than 10 lines, we determined an
appropriate small sample adjustment for the $\sigma$ \citep{keeping}.
In the case of any abundance uncertainty that was calculated to be
less than the uncertainty in the Fe I lines, we conservatively adopted
the value from Fe I for that particular star.  Typically the Fe I
uncertainty is $\sim0.12$ dex.

For those lines with abundances determined via spectral synthesis, we
determined abundance uncertainties based upon the uncertainties
associated with equivalent width measurements.  Continuum placement is
the greatest source of uncertainty, along with the S/N of the region
containing the particular line.  Most of these abundances were
determined from only two lines, thus we calculated the uncertainties
for small samples.  For those elements with only one line, we adopt
the uncertainties determined for the Fe I abundance.

To obtain the systematic uncertainties in the abundances,
we redetermined abundances by individually varying the stellar
parameters by their adopted uncertainties.  We chose a nominal value
of 150\,K for the effective temperature uncertainty, as this value is
similar to the random and systematic uncertainties.
Table~\ref{syserrstab} shows these results.  We find that the
effective temperature contributes most to the abundance uncertainty.
The uncertainty in the surface gravity is somewhat less significant
for most species.  For elements with particularly strong lines,
especially those whose abundances are determined with spectral
synthesis, the microturbulence can be an important source of error.

\begin{deluxetable}{lrrr}
  \tablecolumns{20}
  \tablewidth{0pc}
  \tablecaption{Example Systematic Abundance Uncertainties for HE~0015$-$0048a \label{syserrstab}}
  \tablehead{\colhead{Elem}&\colhead{$\Delta$T$_{eff}$}&\colhead{$\Delta$ log g}&\colhead{$\Delta\xi$}\\
\colhead{}&\colhead{+150\,K}&\colhead{+0.5 dex}&\colhead{+0.3 km/s}}
\startdata
CH    &  0.35 & $-$0.20 &   0.00  \\
Mg I  &  0.11 &  0.06   & $-$0.04 \\
Al I  &  0.17 &  0.18   &    0.11 \\
Si I  &  0.15 & $-$0.05 & $-$0.05 \\
Ca I  &  0.13 &  0.06   &    0.04 \\
Sc II &  0.07 & $-$0.15 & $-$0.05 \\
Ti I  &  0.23 &  0.06   & $-$0.02 \\
Ti II &  0.05 & $-$0.14 & $-$0.08 \\
Fe I  &  0.18 & $-$0.05 & $-$0.05 \\
Fe II &$-$0.01&  0.15   & $-$0.06 \\
Cr I  &  0.21 &  0.08   & $-$0.10 \\
Mn I  &  0.21 &  0.08   & $-$0.14 \\
Co I  &  0.24 &  0.06   & $-$0.11 \\
Ni I  &  0.21 &  0.09   & $-$0.17 \\
Zn I  &  0.00 &  0.00   & $-$0.05 \\  
Sr II &  0.10 &  0.10   & $-$0.20 \\
Ba II &  0.10 &  0.10   & $-$0.07 \\
\enddata
\end{deluxetable}

\subsection{Standard Stars}\label{complit} 
We compared our stellar parameters and [X/Fe] abundances for our four
standard stars (HD~122563, BD~$-$18~5550, CS~22873$-$166, and
CS~22891$-$200) against three studies: \citet{McWilliametal},
\citet{fulbright}, and \citet{cayrel2004}.  We also compared our
derived stellar parameters for CD~$-$38~245 against
\citet{McWilliametal}. The \citet{McWilliametal} study differs from
the other two and this study as those spectra had comparatively low
S/N ($\sim35$).  These stars were chosen from the literature because
they have [Fe/H]$\sim-3$ or below and are in similar evolutionary
stages.  Table~\ref{standardsps} lists the stellar parameters derived in
all three studies, as well as ours, for the five stars and
Table~\ref{sampletbl} includes the derived abundances.

\begin{deluxetable}{lrrrr}
\tablecolumns{21}
\tablewidth{0pc} 
\tablecaption{Literature Values for Stellar Parameters}
\tablehead{\colhead{Study} &\colhead{T$_{eff}$}&\colhead{log g} &\colhead{[Fe/H]}&\colhead{$\xi$}\\
\colhead{} &\colhead{[K]}&\colhead{} &\colhead{}&\colhead{[km/s]}\label{standardsps}}
\startdata
HD~122563\\
This study&  4450 & 0.50  &  $-$2.96 &2.30  \\
Fulbright &  4425 & 0.60  &  $-$2.79 &2.05  \\
McWilliam &\nodata&\nodata&\nodata&\nodata  \\
Cayrel      &  4600 & 1.10  & $-$2.82 & 2.00  \\\hline
BD~$-$18~5550\\
This study&4600  &0.80   &$-$3.20&1.70     \\
Fulbright &\nodata&\nodata&\nodata&\nodata \\
McWilliam &4790   &1.15   &$-$2.91&2.14    \\
Cayrel    &4750   &1.40   &$-$3.06&1.80    \\\hline
CD~$-$38~245\\
This study&4560   &0.95   &$-$4.35&2.15    \\
Fulbright &\nodata&\nodata&\nodata&\nodata \\
McWilliam &4730   &1.80   &$-$4.01&1.97    \\
Cayrel    &4800   &1.50   &$-$4.19&2.20    \\\hline
CS~22873$-$166\\
This study&4375   &0.20   &$-$3.14& 2.60   \\
Fulbright &\nodata&\nodata&\nodata&\nodata \\
McWilliam &4480   &0.80   &$-$2.90& 3.01   \\
Cayrel    &4550   &0.90   &$-$2.97& 2.10   \\\hline 
CS~22891$-$200\\
This study&4500   & 0.45   & $-$3.92  & 2.60 \\ 
Fulbright &\nodata& \nodata& \nodata& \nodata\\
McWilliam &4700   & 1.00   & $-$3.49  & 2.51 \\ 
Cayrel    &\nodata& \nodata& \nodata& \nodata\\
\enddata
\tablecomments{References: \citet{fulbright}, \citet{McWilliametal}, \citet{cayrel2004}}

\end{deluxetable}

The \citet{McWilliametal} study contains four stars (including
CD~$-$38~245) which overlap with our study.  That work utilized model
atmospheres from \citet{kurucz} with MOOG.  We use an updated version
of this code, which does explicitly deal with Rayleigh scattering as a
continuum opacity source; see paragraph 1 of Subsection~\ref{analtech}.
The effective temperatures for this study were derived from
photometry, with the microturbulence determined from Fe I lines and
the surface gravity determined from the ionization balance of Fe I and
Fe II lines.  We find that our temperatures are lower by $\sim170$
\,K.  Our derived surface gravities are also lower, as a result of the
lower temperatures.  We also find a $\sim-0.3$ dex offset in the
[Fe/H] values.  The cause of these discrepancies is likely twofold:
the different temperature scales and how the most recent version of
MOOG explicitly deals with the Rayleigh scattering opacity (see
Table~\ref{2002comp}).  Both lead to lower temperatures, surface
gravities, and abundances.

Despite these absolute offsets, the derived abundance ratios have only
small offsets, with the average offset between the two studies in
$\langle$$\Delta$[X/Fe]$\rangle$ for all three stars is 0.04$\pm$0.24 dex, where
$\Delta$[X/Fe] is [X/Fe]$_{\rm{Standard}}$$-$[X/Fe]$_{\rm{MIKE}}$.
For all three stars, Al and Si had the largest offsets, where
\citet{McWilliametal} derive systematically higher abundances.  The Al
abundances for \citet{McWilliametal} may generally be high, as they
were found to be higher ($\Delta$[Al/Fe] $\sim0.5$ dex) than
\citet{gs88}, with no conclusion for the cause of the offset being
reached.  The Al and Si lines are located in low S/N regions in both
studies; however, the measurement of the Al lines is difficult due to
the suppression of the continuum from Ca H and K lines.  As stated in
\citet{al_nlte}, even under the assumption of LTE, small changes in
the effective temperature and surface gravity greatly change the line
profiles.  In addition, the proximity of the H$\delta$ line produces
larger uncertainties in our Si measurement, even when the feature is
taken into account in the synthetic spectra.  These reasons would
produce a larger scatter in the measurements of Al and Si lines,
compared to the other elements, and may explain the derived abundance
discrepancy. When these two elements are removed from consideration
the $\langle\Delta$[X/Fe]$\rangle$ becomes $-0.02\pm0.13$ dex.

The only star which overlaps with the \citet{fulbright} study is
HD~122563, though this particular study was chosen because the stellar
parameters were determined spectroscopically, in the same manner as
the stellar parameters determined from the MIKE spectra.
\citet{fulbright} used \citet{kurucz} model atmospheres with MOOG as
well.  We found that the effective temperatures, surface gravities,
and microturbulence values agree within the uncertainties associated
with both studies, while $\Delta$[Fe/H] is $-0.17$ dex.  Similar to
the effects seen in comparison with the \citet{McWilliametal} study,
this is due to the fact that the version of MOOG used in the
\citet{fulbright} study did not explicitly handle the calculation of
scattering from pure absorption in terms of the continuum opacity.

The derived abundance ratios are in good agreement, with
$\langle$$\Delta$[X/Fe]$\rangle=0.02\pm0.09$ dex.  The largest
discrepancy lies with the [Mg/Fe] ratio, with $\Delta$[Mg/Fe] = 0.20
dex, which was derived with a different set of lines and different
oscillator strengths between this study and the \citet{fulbright}
study.

The \citet{cayrel2004} study has four stars in common with ours
(including CD~$-$38~245).  This study uses OSMARCS atmospheric models
with the LTE synthetic spectrum code ``turbospectrum'', which does
explicitly handle the calculation of the scattering contribution with
regard to the continuum opacity.  The effective temperatures were
determined via (V$-$K) colors, which leads to higher temperatures
($\sim200$\,K) and surface gravities ($\sim0.6$ dex) compared to our
spectroscopically derived stellar parameters.  We ran the measured
equivalent widths of \citet{cayrel2004} for BD~$-$18~5550 through
Cashcode and found that our derived stellar parameters (4560\,K, 0.6
dex, $-3.22$, and 1.7 km/s in effective temperature, surface
gravity, [Fe/H], and microturbulence, respectively) were in agreement
($\Delta$[Fe/H]$=0.02$ dex).

In comparing the [X/Fe] values between our analysis of the
\citet{cayrel2004} equivalent widths and theirs, we find that our
spectroscopically derived stellar parameters also lead to higher
[Mg/Fe] values by $\sim0.25$ dex due to the gravity sensitive nature
of the Mg lines.

The derived abundance ratios also agree well, with
$\langle$$\Delta$[X/Fe]$\rangle=-0.03\pm0.14$ dex.  The largest sources of
discrepancy for all stars are Al and Si, in addition to Mg.  In
CS~22873$-$166, there is a 0.45 discrepancy between the [Sr/Fe]
values.  The S/N near both lines in the MIKE spectrum is $\sim20$.
Additionally, the lines are very strong, making continuum placement a
large source of uncertainty, as minor adjustments to the continuum
level result in large changes in the abundance.

\section{Abundance Results and Discussion}\label{interp}
Table~\ref{sampletbl} lists abundance results derived from the MIKE
spectra.  Figures~\ref{cashrel} and~\ref{mikerel} include the [X/Fe]
abundance ratios derived from the MIKE spectra plotted against [Fe/H]
for all stars in the sample.  These abundance ratios are overplotted
against the \citet{cayrel2004} abundances as a point of comparison.
The table also includes the abundances derived from the HRS spectra,
though we do not discuss these further.  These abundances will later
be included in the full $\sim500$ star CASH sample. For each element,
Table~\ref{results} lists the parameters of a least squares linear
trend versus metallicity, the abundance scatter, and the average
[X/Fe] value, if applicable.

\begin{deluxetable}{lrrll}
  \tablecolumns{20}
  \tablewidth{0pc}
  \tablecaption{Summary of Abundance Trends\label{results}}
  \tablehead{\colhead{Elem}&\colhead{$\langle$[X/Fe]$\rangle$}&\colhead{slope vs [Fe/H]}&\colhead{$\sigma$}&\colhead{n$_{stars}$}}
\startdata
CH    &    0.02\tablenotemark{a}&$-0.52$\tablenotemark{b}&0.41 &20\\
Mg I  &    0.56    & \nodata &0.14 & 20  \\
Al I  &    \nodata & 0.11    &0.22 & 19  \\
Si I  &    \nodata &$-$0.20  &0.16 & 19  \\
Ca I  &    0.42    & \nodata &0.11 & 20  \\
Ti II &    0.22    & \nodata &0.16 & 20  \\
Sc II &    \nodata & \nodata &0.19 & 20  \\
Cr I  &    \nodata & 0.26    &0.02 & 19  \\
Mn I  &    $-$0.49 & 0.24    &0.09 & 19  \\
Co I  &    0.42    & \nodata &0.11 & 20  \\
Ni I  &    0.05    & \nodata &0.15 & 20  \\
Zn I  &    \nodata & $-$0.42 &0.04 & 19  \\
Sr II &   $-$0.66  & \nodata &0.57 & 20  \\
Y  II &   $-$0.46  & \nodata &0.32 & 15  \\
Zr II &   $-$0.08  & \nodata &0.26 & 11  \\
Ba II &    $-$0.95 & \nodata &0.25 & 20  \\
\enddata
\tablenotetext{a}{Average [C/Fe] value calculated for non-CEMP stars.}
\tablenotetext{b}{Slope calculated for log(L/L$_{\odot}$) vs [C/Fe].  See Figure~\ref{cab}.}
\end{deluxetable}

\subsection{Light Elements}\label{int_lightel}
\begin{figure}[]
\includegraphics[width=0.5\textwidth]{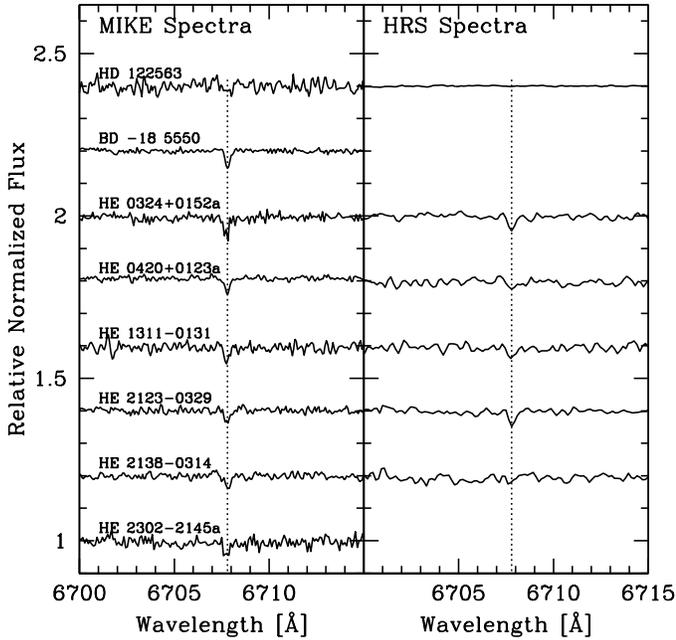}
\caption{Li $\lambda$6707 line detections in MIKE (black solid line)
  and CASH (red dotted line).  Plotted (dashed line) is the location
  of the feature.  Also included for comparison is HD~122563, a star
  for which no Li was detected.\label{lidet}}
\end{figure}

\begin{figure}[]
\includegraphics[width=0.5\textwidth]{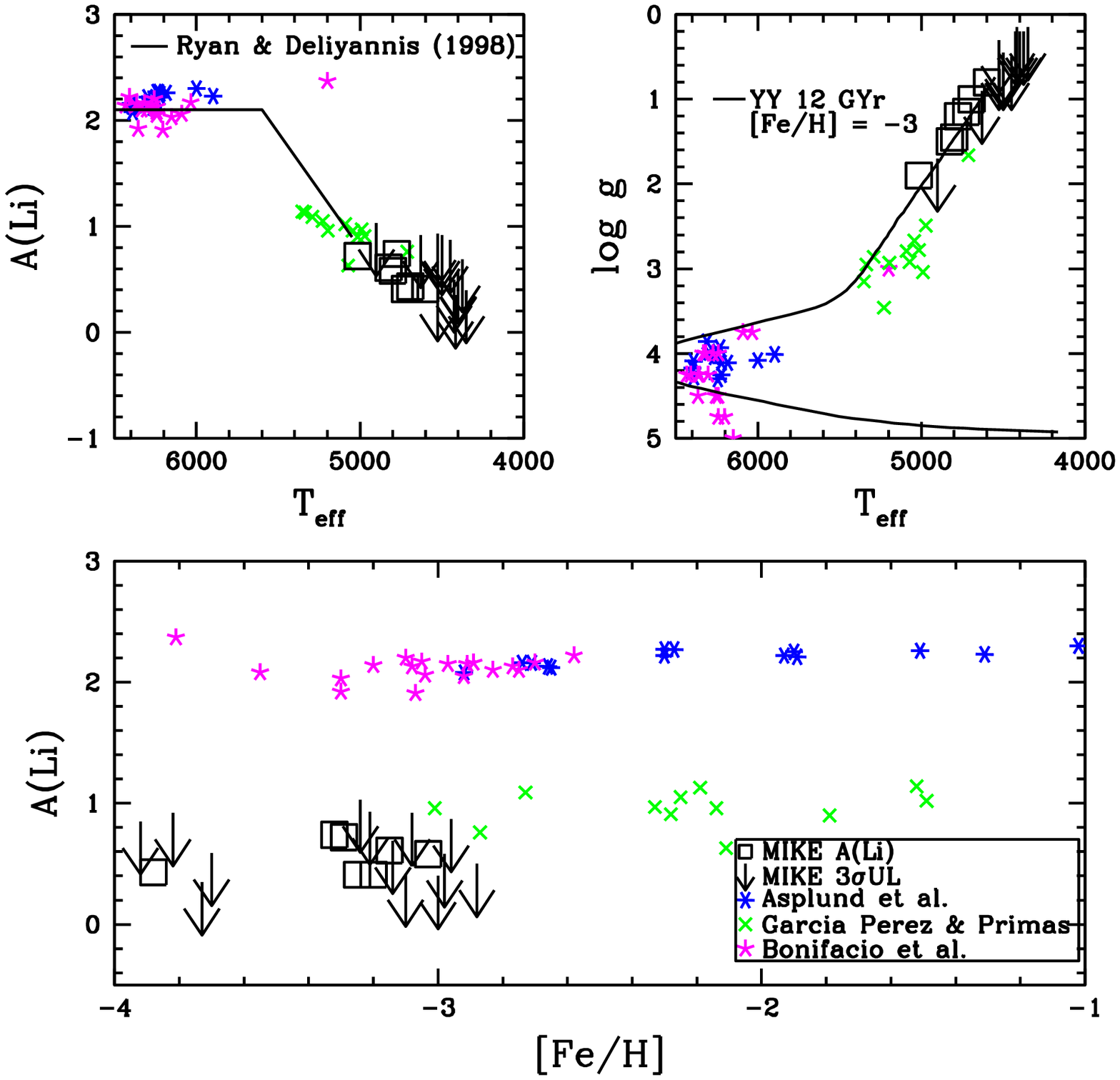}
\caption{MIKE Li abundances (open squares) and upper limits (arrows)
  plotted against [Fe/H] along with Li abundances from
  \citet{asplund06}, \citet{garciaperez_primas2006}, and
  \citet{bonifacio} (bottom), and effective temperature with the
  expected Li dilution curve from \citet{ryan1998} (top left)
  overlayed.  Effective temperatures and surface gravities are plotted,
  along with a [Fe/H]=$-$3, 12 Gyr Yale-Yonsei isochrone
  \citep{green,Y2_iso} for each star plotted in the bottom panel (top
  right).
\label{liab}}
\end{figure}

We detected Li in six of our stars: HE~0324$+$0152a, HE~0420+0123a,
HE~1311$-$0131, HE~2123$-$0329, HE~2138$-$0314, HE~2302$-$0849a, and
in one standard star, BD~$-$18~5550.  These detections were seen in
both the snapshot and high-resolution spectra for HE~0420+0123a,
HE~1311$-$0131, HE~2123$-$0329, and HE~2138$-$0314. BD~$-$18~5550 and
HE~2302$-$0849 do not have a corresponding snapshot spectrum.  In
Figure~\ref{lidet} we show the Li line in the HRS and MIKE data for
all these stars.  Otherwise, we determined upper limits. In each case
the S/N of the region was measured and the corresponding 3$\sigma$
upper limit on the equivalent width was calculated following
\citet{bohlin} and \citet{frebel_he1300}.  Figure~\ref{liab} shows the
Li abundance as a function of metallicity and effective temperature,
along with the HR diagram of the stars with detected Li.

The Spite plateau \citep{spite_lithium_pl82} is an observational
discovery that describes a constant Li abundance in low-metallicity
stars near the main-sequence turn-off.  Metal-poor stars near or on
the main sequence have not yet burned their surface Li; it is
therefore thought that the stars that populate the Spite plateau can
be used to infer details about the primordial Li abundance.  As stars
evolve off the main sequence and up the red giant branch (RGB), their
convection zone deepens and the atmospheric Li abundance is depleted
through burning and convective dredge up.  Any enhancement in the Li
abundance (e.g.,~\citealt{cash1}) is likely from some form of Li
synthesis that occurs during the course of stellar evolution, perhaps
due to the Cameron-Fowler mechanism \citep{cameronfowler}.  All of the
stars in the pilot sample are on the giant branch and, thus, are
expected to have depleted Li abundances.  Due to the evolutionary
status of the pilot sample, we cannot comment on the nature of the
Spite plateau.  Accurate Li abundances require great care in the
effective temperature determination, as the Li abundance is extremely
temperature sensitive.  Most Li abundance studies
(e.g.,~\citealt{RBDT96,asplund06,garciaperez_primas2006,bonifacio,melcas})
employ long baseline (e.g.,~V$-$K) photometric effective temperatures.
Keeping the different temperature scales in mind, we find that our
abundances qualitatively fall along an extrapolation of the
\citet{ryan1998} Li dilution curve.

\begin{figure}[]
\includegraphics[width=0.5\textwidth]{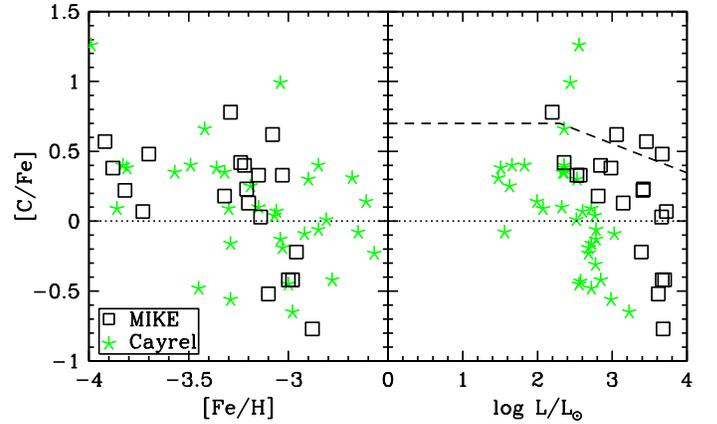}
\caption{\label{cab}[C/Fe] abundance ratios plotted against [Fe/H]
  (left) and luminosity along with the CEMP defining line, which
  changes over the course of the stellar lifetime
  \citep{aoki_cemp_2007} (right), along with the calculated
  \citet{cayrel2004} C abundances.  The [C/Fe] abundances clearly
  decline as stars ascend the giant branch.}
\end{figure}

\begin{figure}[]
\includegraphics[width=0.5\textwidth]{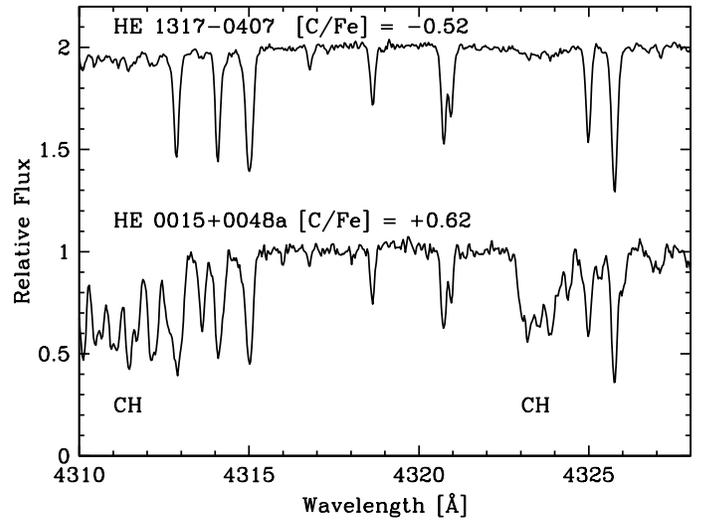}
\caption{\label{ch} HRS spectra of a CEMP star (HE~0015+0048a) and a
  non-CEMP (HE~1317$-$0407) star of similar temperatures ($\sim 4550
  $\,K) and metallicities ($\sim -3.1$).}
\end{figure}

Figure~\ref{cab} shows a plot of our derived [C/Fe] abundances
against [Fe/H] and luminosity.  The \citet{cayrel2004} abundances are
also shown for reference.  The abundance offset as a function of
luminosity between the two samples is due to the different temperature
scales used, as also quantified in Table~\ref{syserrstab}.  Generally,
we find a large spread in the [C/Fe] abundance ratios, from
$\sim-0.80$ to $\sim 0.8$ dex, with $\sigma$ = 0.41 dex.  In
Figure~\ref{ch}, we show spectra for two stars of similar temperature
($\sim4550$\,K) that differ in [C/Fe] by $\sim1.0$ dex.

However, the interpretation of the observed [C/Fe] ratio must be
carefully evaluated.  As stars evolve up the RGB, the C abundance
drops due to CN cycling and convective dredge up. Thus, we expect that
the observed C abundances derived for our stars should be lower than their
initial abundances.  \citet{gratton2000} found that metal-poor field
stars on the upper RGB had [C/Fe] ratios that were $\sim 0.5$ dex
lower than metal-poor field stars on on the main sequence or turn-off.
We very clearly see this trend in the pilot sample stars, all of which
are on the RGB, in Figure~\ref{cab}, where the [C/Fe] ratio drops by
$\sim 1.2$ dex over one solar luminosity.

\citet{aoki_cemp_2007} provided a new definition for CEMP star status:
stars with log(L/L$_{\odot}$) $< \sim2.3$ and [C/Fe] $\geq$ 0.70 
and stars with log(L/L$_{\odot}$) $> \sim2.3$ and [C/Fe] $>$
(3.0$-$log(L/L$_{\odot}$)) are considered CEMP stars (see the right
panel of Figure~\ref{cab}).  This definition takes into account the
decrease in the surface C abundances as a function of evolutionary
status.  For our sample, the definition would indicate that 4 stars
are significantly enhanced in their [C/Fe] ratios compared to the rest
of the sample.  This is still true when the effects on the abundances
due to the use of different temperature scales is taken into account.
In order to study the initial stellar abundances, corrections, up to
$\sim1$ dex for the highest luminosity stars, should be applied.
Thus, if we were to correct for this depletion, one can estimate that
the ISM from which these stars formed was enriched in C by a factor of
$\sim 10$ or more.

Nevertheless, the large scatter in the [C/Fe] ratios of metal-poor
stars indicates a complex production history of C.  Additionally, the
sizable fraction of metal-poor stars that show enhancement in their
[C/Fe] ratio \citep{frebel_bmps,cohen2006,carolloCEMP}
make it important to study the origin of C in the early universe.  For
metal-poor stars, the most important production site of C is massive
stars that can release CNO elements during the course of stellar
evolution through various supernova outbursts
(e.g.,~\citealt{fryer2001,meynet06,Kobayashi2011}).

With these considerations in mind, our C abundances suggest one of the
following three scenarios: i) strong stellar winds from massive stars
released C from their atmospheres, enriching only the local ISM, ii)
after enrichment from the supernovae of massive stars and/or stellar
winds in the early universe, there was only inhomogeneous mixing of
the ISM, or iii) the sample stars did not all form from the same
molecular cloud or in the same host system.  For example, some of the
surviving ultra-faint dwarf spheroidal galaxies show a $\sim1$ dex
spread in their C abundances \citep{Norris2010_1,Norris2010_2}.
Scenario two does not agree with the small scatter seen in their
$\alpha$-element ratios.  The first and third scenarios, however, are
not mutually exclusive.  Future modeling of the nucleosynthetic yields
of massive Population III stars will facilitate a better understanding
of early carbon production.  Given the size and metallicity range of
the pilot sample, unfortunately nothing can be said about the CEMP
frequency, though we will be able to evaluate this using the full CASH
sample.

Finally, fine-structure lines of C and O are thought to play a role in
the transition from Population III to Population II stars through the
cooling of gas clouds of the early universe \citep{brommnature}.  This
hypothesis can be tested by comparing the abundances of C and O of
metal-poor stars to an abundance transition discriminant as described
in \citet{dtrans}.  Stars with [Fe/H]$<-3.5$ are particularly
interesting in this regard; however, all stars of the pilot sample are
too C-rich to indicate that fine structure cooling \textit{did not}
play a role.

\subsection{$\alpha$-Elements}
The $\alpha$-elements are created both in core-collapse supernova and
through stellar nucleosynthesis in high-mass stars.  While the
dominant isotope of Ti is not technically an $\alpha$-element, it
shows a similar abundance pattern to the $\alpha$-elements and thus is
included here (e.g.,~\citealt{woosley_weaver_1995}).  Magnesium and
Ca, as well as the other $\alpha$-elements, have been shown to be
overabundant, compared to the Solar System [$\alpha$/Fe] ratio, at low
metallicities in field halo stars at the $\sim0.4$ dex level (e.g.,~
\citealt{mcwilliamreview}, \citealt{cayrel2004}, \citealt{frebel10}).
This has been explained by the occurrence of core-collapse supernovae
in the early universe, which produce an overabundance of the
$\alpha$-elements compared to Fe.  Later generations of supernovae,
specifically Type Ia, produce more Fe, driving down the [$\alpha$/Fe]
ratio to what we see today in the Sun and similar young, metal-rich
stars.  This downturn in the [$\alpha$/Fe] ratio occurs at [Fe/H]
$\sim-1.5$.  As our sample does not reach [Fe/H]$ > -2.0$, we do not
expect to see this downturn in the [$\alpha$/Fe] abundance ratios.

As seen in other halo stars, the [Mg/Fe] ratio is enhanced relative to
the solar ratios in all our sample stars at 0.56 dex.  This is also
seen in the \citet{cayrel2004} stars as well, though there is an
offset of $\sim 0.25$ dex between these two samples, with ours having
the larger value.  This is due to the differences in the effective
temperature scales chosen.  The \citet{cayrel2004} study used (higher)
photometric temperatures.  To demonstrate this, we took the
\citet{cayrel2004} equivalent widths for BD~$-$18~5550 and ran them
through the Cashcode pipeline.  We obtained a temperature different by
150\,K, resulting in an offset of 0.5 dex in the surface gravity.  In
most elemental ratios these effects cancel, but Mg, especially the
triplet, is gravity sensitive, resulting in a $\sim 0.25$ dex offset
in [Mg/Fe] between the two studies.

The [Ca/Fe] ratio is also enhanced in our sample stars relative to the
solar ratio at 0.42 dex.  The [Ti/Fe] ratio is found to be enhanced
relative to the solar ratio in all but two stars, HE~2148$-$1105a and
HE~2302$-$2145a, with [Ti/Fe] values of $-$0.14 and $-$0.03
respectively; however, these stars both show the expected enhancement
in [Ca/Fe] (0.17 and 0.31).  The [Mg/Fe] values (0.27 and 0.30) in
these stars is somewhat lower compared to the values in our other
stars.  The average [Ti/Fe] value for this study is 0.22.  The average
[$\alpha$/Fe] value is $\sim0.4$ dex and thus corresponds well to the
average halo [$\alpha$/Fe] values.

\subsection{Fe-Peak Elements}
The [Sc/Fe] ratio for the stars in the sample is generally clustered
around the solar abundance ratio.  The [Cr/Fe] and [Mn/Fe] abundance
ratios for all our sample stars are found to be deficient relative to
the solar abundance by $-0.23$ and $-0.49$ dex, respectively. We
remind the reader that the [Cr/Fe] ratios are based upon only the Cr I
abundances, as there is a $\sim 0.35$ dex offset between Cr I and Cr
II derived abundances.  The [Co/Fe], [Ni/Fe], and [Zn/Fe] ratios in
the sample stars are generally enhanced relative to the solar
abundance ratios by 0.42, 0.05, and 0.25 dex, respectively.

The Fe-peak elements are created in various late burning stages (see
\citealt{woosley_weaver_1995}), as well as in supernovae. Our Fe-peak
abundance trends follow those of other halo star samples and generally
indicate a successive increase of these elements over time
(e.g.,~\citealt{mcwilliamreview}).  We will use the Fe-peak elemental
abundances of the large CASH sample to put constraints on the
nucleosynthesis yields of the progenitor stars.  For example, the
measured Zn abundances, which we can measure in the HRS spectra, are
sensitive to the explosion energy of supernovae
\citep{nomoto2006,heger_woosley2010}.

\subsection{Neutron-Capture Enhanced Stars}
The study of neutron-capture elements allows for testing of different
sites of nucleosynthesis, beyond proton and $\alpha$-capture.  See
\citet{sneden_araa} for a comprehensive overview of the
neutron-capture stellar abundances.  Neutron capture occurs mainly in
two locations: in the envelopes of highly evolved asymptotic giant
branch (AGB) stars (s-process) and in some sort of explosive event,
likely a core-collapse supernova (r-process).  The contributions of
each process to the total elemental abundance of the neutron-capture
elements in a given star can be determined by evaluating the
collective neutron-capture abundance patterns.  For example, in the
Solar System abundances, the s-process contributes $\sim80\%$ of the
Ba abundance, with a $\sim20\%$ contribution from the r-process,
whereas Eu is made almost entirely from the r-process
\citep{sneden_araa}; however, these ratios may differ in the early
universe.  Unfortunately, there are not enough neutron-capture
elements detectable in the HRS snapshot spectra to determine whether
the abundances of neutron-capture elements in a given star have an s-
or r-process origin.

\begin{figure}[]
\includegraphics[width=0.5\textwidth]{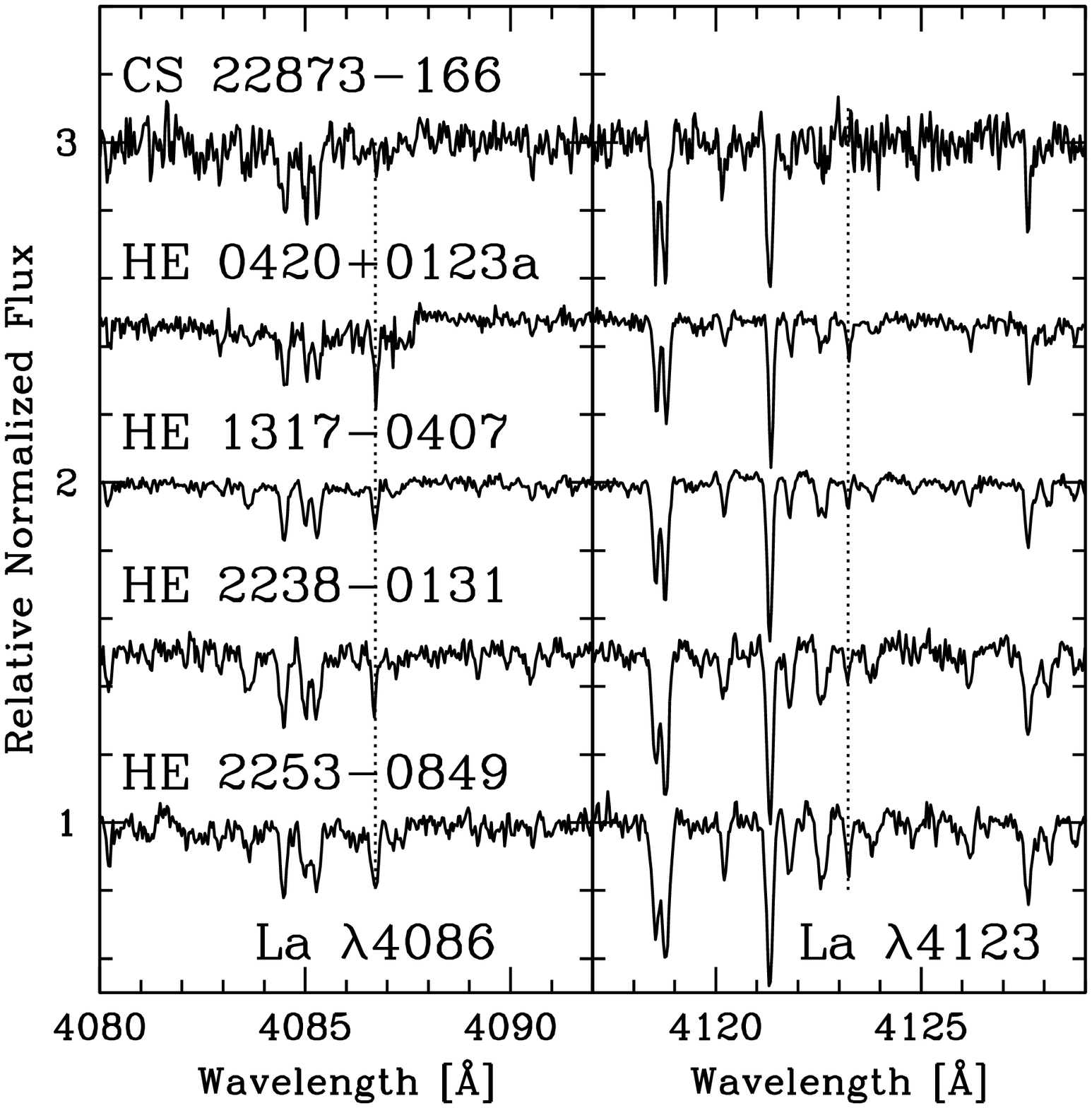}
\caption{\label{ladet} La $\lambda$4086 and $\lambda$4123 line
  detections in MIKE (black solid line) spectra.  Plotted in each
  panel (dashed line) is the location of the features.  For
  CS~22873$-$166, only the $\lambda4086$ line is a detection.}
\end{figure}

\begin{figure}[]
\includegraphics[width=0.5\textwidth]{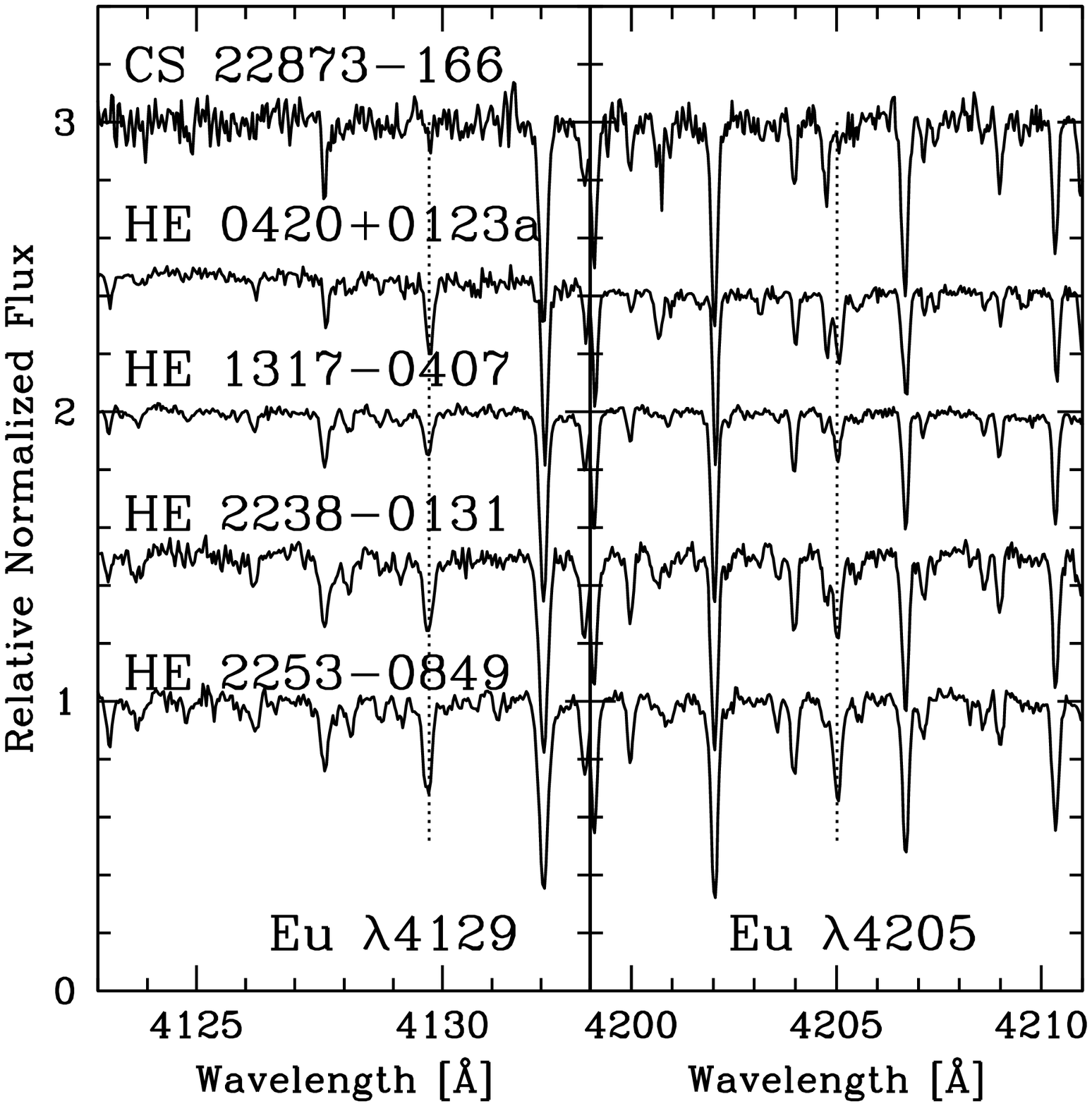}
\caption{\label{eudet} Eu $\lambda$4129 and $\lambda$4205 line
  detections in MIKE (black solid line) spectra.  Plotted in each
  panel (solid dashed) is the location of the features. For
  CS~22873$-$166, only the $\lambda4129$ line is a detection.}
\end{figure}

We measured Sr and Ba for all stars in the MIKE sample.  We also
detected Y and Zr in many of the stars as well, though the S/N near
the Y lines often preclude us from measuring it, and the Zr feature is
often too weak.  We measured La in only four pilot sample stars:
HE~2238$-$0131, HE~0420$+$0123a, HE~1317$-$0407, and HE~2253$-$0840
and in the standard star CS~22873$-$166.  Europium lines have been
detected in the high-resolution spectra of the same five stars.  In
Figures~\ref{ladet} and~\ref{eudet} we show line detections of La and
Eu in each star, respectively.  The four sample stars are all
neutron-capture enhanced, ranging in [Eu/Fe] from 0.02 to 0.79 dex,
while CS~22873$-$166 is depleted in [Eu/Fe] relative to the solar
abundance ratio.  According to \citet{heresI} HE~0420+0123a is a
mildly r-process enhanced (r-I) star due to its [Eu/Fe] ratio (0.79)
and Ba/Eu ratio ($-0.71$).

\begin{figure}[]
\includegraphics[width=0.5\textwidth]{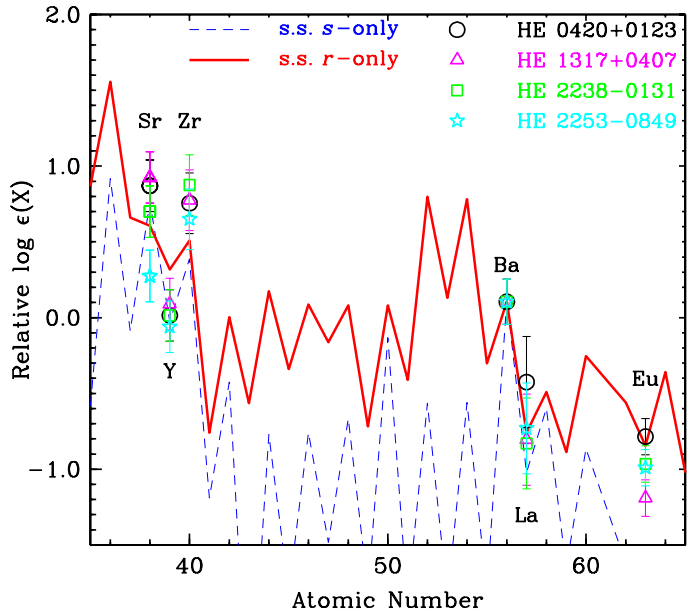}
\caption{\label{rproc}Relative log $\epsilon$(X) abundances for all
r-process enhanced stars in the MIKE sample.  Abundances are adjusted
to the Ba abundance to fit a scaled Solar System r-process curve (red
line).  The Solar System s-process curve is also plotted (blue dashed
line).}
\end{figure}

For each of these neutron-capture enhanced stars, we normalized their
log $\epsilon$(X) abundances relative to the Ba abundance and plotted
them against a Solar System r-process abundance pattern.  We also
normalized the Solar System s-process abundance pattern to fit the
derived Ba abundance.  We found that the r-process pattern seemed to
fit the ratio of La to Eu better than the s-process pattern, as the Eu
abundances would have to be an order of magnitude lower to match the
s-process pattern.  Figure~\ref{rproc} shows this analysis.  The first
neutron-capture peak elements (Sr, Y, Zr) show a $\sim 0.3$ dex range
of abundances, though some are enhanced relative to the r-process
curve.  This may indicate two things: i) the r-process seems a
likely explanation for the Eu abundances and ii) a nucleosynthetic
event besides the same r-process that formed a portion of the
neutron-capture elements in the Solar System must be contributing in
some of the stars to explain their first neutron-capture peak
elemental abundance range.

\subsection{Stars with [Fe/H]$ < -3.5$}
Four stars in the pilot sample have [Fe/H]$ < -3.5$, along with an
additional standard star, CS~22891$-$200, which was found to have a
lower [Fe/H] value ($-3.92$) than was previously published by
\citealt{McWilliametal} ($-3.49$).

All these extreme EMP (EEMP) stars have enhancements in the
[$\alpha$/Fe] ratios, depletion in some of the Fe-peak abundance
ratios, and none are neutron-capture enhanced.  We detected the
$\lambda$6707 Li I line in HE~2302$-$2154a, which can be expected
given that it is the warmest of the EEMP stars at T$_{\rm{eff}}$ =
4675\,K.  Generally, these stars are indistinguishable from the rest of
the sample, with the following exception.

It is noteworthy that all five EEMP stars show enhancement in their
[C/Fe] abundance ratios.  This confirms the trend toward higher [C/Fe]
ratios towards lower values of [Fe/H], noted first by
\citet{rossi2005} and \citet{lucatello2006}.  Without consideration
for the evolutionary stage of the stars, the average [C/Fe] ratio for
the EEMP stars is 0.43, while the rest of the sample has an average
[C/Fe] ratio of 0.07 dex.  However, given the high luminosity of the
EEMP stars, the [C/Fe] ratios have been depleted due to the operation
of the CN cycle.  Taking this effect into account, two stars are
mildly C enriched and two are considered CEMP stars following the
definition of \citet{aoki_cemp_2007}.  The two CEMP stars lack
neutron-capture enhancement, which indicates that these stars are
CEMP-no stars \citep{bc05}, adding to the growing number of these
objects at the lowest metallicities.

We did not detect either the La or Eu lines in any of these stars.
This is largely due to the fact that it is difficult to detect Eu in
EMP stars unless the Eu abundance is significantly enhanced.  Given
that we do not see strong La enhancement along with the carbon
enhancement in the EEMP stars, it is likely that their observed
abundance patterns were the result of massive stars, rather than low
mass stars, which produce La in the AGB stage of
stellar evolution.

\subsection{Binary Fraction}
For twelve of the pilot sample stars, we derived multiple-epoch
measurements of their radial velocity.  For three stars, we have
additional radial velocity measurements from the solar
spectrum-contaminated HET HRS spectra.  For the remaining two stars,
we do not have multiple measurements.  Generally, the radial velocity
measurements were taken with a baseline of at least one year and in
some cases, three years.  The HRS and MIKE radial velocities agree to
within 2.5 km/s for those stars.  We find one binary candidate in our
sample.  HE~0015+0048a is the one exception; the time span between
measurements is three years and we find a radial velocity variation of
8.0 km/s.  Future monitoring of this star will confirm its
binarity. This sample is too small to make any definitive statement on
the binary fraction of metal-poor stars.

\section{Summary}\label{conclu}
We have presented the abundances or upper limits of 20 elements (Li,
C, Mg, Al, Si, Ca, Sc, Ti, Cr, Mn, Fe, Co, Ni, Zn, Sr, Y, Zr, Ba, La,
and Eu) for 16 new stars and four standard stars derived from
high-resolution, high S/N MIKE spectra via traditional manual analysis
methods using the MOOG code.  We find that, with the exception of Mg,
our abundances match well with those of other halo stars reported in
the literature, e.g., \citet{cayrel2004}.

In the pilot sample we find several distinct chemical groupings of
stars, indicating different enrichment mechanisms may apply for each
of these groups, though the exact mechanism is still uncertain.
We find four new stars with $\mbox{[Fe/H]} < -3.6$, where the
metallicity distribution function severely drops.  We find
CS~22891$-$200 to have a lower $\mbox{[Fe/H]}$ value than reported by
\citet{McWilliametal}, bringing the total number of $\mbox{[Fe/H]} <
-3.6$ stars to five.  All of these stars are enhanced in their [C/Fe]
abundance ratios relative to the solar values.  We have four CEMP
stars in our sample. Two of these are EEMP CEMP-no stars, which
confirms the trend of an increasing enhancement in C towards lowest
metallicities without invoking the contribution of AGB stars at the
lowest metallicities.  This may suggest that massive stars released C
from their atmospheres and enriched the local ISM and/or that the
sample stars did not all form in the same region.  We detected La and
Eu in five stars.  Of these, four are neutron-capture element enhanced
with r-process signatures, based upon their [Eu/Fe] ratios, ranging
from 0.02 to 0.79; the star with the highest [Eu/Fe] abundance ratio
is an r-I star.  For the remaining stars, we generally find scaled
Solar System abundance ratios with very small scatter in the
abundances.  This may indicate the presence of core-collapse
supernovae in the early universe.  In the pilot sample, we have one
star that is a binary candidate.  Future monitoring of this star will
determine its binarity status.

We presented a calibration of the Cashcode pipeline.  We find
agreement between the spectroscopic stellar parameters derived from
the high-resolution spectra and the snapshot spectra to within $\Delta
T_{eff}\pm 55$\,K, $\Delta$ log g$\pm$ 0.3 dex,
$\Delta\mbox{[Fe/H]}\pm 0.15$ dex, and $\Delta\xi\pm 0.21$ km/s.
These fall within the expected uncertainties associated with
snapshot-quality data.  We also find that the abundances derived from
the HRS spectra using the pipeline are in agreement to within 1.5
$\sigma$.  The Cashcode pipeline will be employed for the full
$\sim500$ star snapshot sample.  This sample will be used to determine
carbon and neutron-capture enhancement frequencies, to better
understand supernova nucleosynthesis and early universe chemical
enrichment processes, and find new astrophysically interesting stars
that merit further study.  One star has already been identified as
neutron-capture enhanced in our sample, based on its high Eu abundance
and another separate star has also been singled out.  The latter star
is a CEMP star, with r-process and s-process elemental abundance
enhancements.  Both of these will be further analyzed in a later
paper as part of this series.

\acknowledgments

The Hobby-Eberly Telescope (HET) is a joint project of the University
of Texas at Austin, the Pennsylvania State University, Stanford
University, Ludwig-Maximilians-Universität München, and
Georg-August-Universität Göttingen. The HET is named in honor of its
principal benefactors, William P. Hobby and Robert E. Eberly.  We are
grateful to the Hobby-Eberly staff for their assistance in obtaining
the data collected for this paper.  We thank John Norris and Norbert
Christlieb for their valuable contributions to the discovery of the
bright metal-poor stars analyzed here.  J.K.H. and T.C.B. acknowledge
partial support through grants PHY 02-16783 and PHY 08-22648: Physics
Frontier Center/Joint Institute for Nuclear Astrophysics (JINA).
A.F. acknowledges support of a Clay Fellowship administered by the
Smithsonian Astrophysical Observatory.  I.U.R. is supported by the
Carnegie Institution of Washington through the Carnegie Observatories
Fellowship. C.S. is supported through NSF grant AST-0908978.

\end{document}